\documentclass[
notitlepage,
amsmath,amssymb,
floatfix,
]{revtex4-1}
\usepackage{color}
\usepackage{xcolor}
\usepackage{graphicx}
\usepackage{dcolumn}
\usepackage{bm}
\usepackage[colorlinks=true,urlcolor=blue,citecolor=blue,linkcolor=blue]{hyperref}

\begin{document}

\title{Optimized Planar Microwave Antenna for Nitrogen Vacancy \\ Center based Sensing Applications}

\author{Oliver Roman Opaluch}
\author{Nimba Oshnik}
\altaffiliation[Present address: ]{Universit{\"a}t des Saarlandes, Faculty of Natural Sciences and Technology, Physics, Campus E2.6, D-66123 Saarbr{\"u}cken, Germany}
\author{Richard Nelz}
\author{Elke Neu}
 \email{nruffing@rhrk.uni-kl.de}
\affiliation{Department of Physics, University of Kaiserslautern, Erwin-Schrödinger-Straße 67663 Kaiserslautern}

\date{\today}

\begin{abstract}
Individual nitrogen vacancy (NV) color centers in diamond are versatile, spin-based quantum sensors. Coherently controlling the spin of NV centers using microwaves in a typical frequency range between 2.5 and 3.5 GHz is necessary for sensing applications. In this work, we present a stripline-based, planar, $\Omega$-shaped microwave antenna that enables to reliably manipulate NV spins. We find an optimal antenna design using finite integral simulations. We fabricate our antennas on low-cost, transparent glass substrate. We demonstrate highly uniform microwave fields in areas of roughly 400 $\times$ 400 $\mu$m$^2$ while realizing high Rabi frequencies of up to 10 MHz in an ensemble of NV centers.
\end{abstract}
\maketitle

\section{Introduction}
\label{Sec1}
 Nitrogen vacancy (NV) centers in diamond consist of a substitutional nitrogen atom and an adjacent lattice vacancy. In recent years, this crystal defect in its negatively charged form NV$^-$ has been extensively studied as a versatile spin-based quantum sensor (for simplicity, we use the term NV center for the negatively charged state throughout this manuscript). The NV center forms an electronic S = 1 spin system with a zero field splitting (ZFS) of 2.87 GHz between the $m_s$ = 0 and the degenerate $m_s$ = $\pm$ 1 ground state. Manipulating these highly-coherent spin states is accomplished using microwaves (MW) e.g. via coherent oscillations (Rabi oscillations). These form the basis of spin-based quantum sensing using NV centers. An external magnetic field parallel to the NV's high symmetry axis lifts the degeneracy between $m_s$ = $\pm$ 1 states via the Zeeman effect, whereas a perpendicular field induces spin mixing. In addition, mechanical pressure, temperature and electrical fields interact with the NV spin, consequently making it a versatile atomic scale sensor \cite{jaskula2017superresolution,barson2017nanomechanical,pham2016nmr,arai2015fourier,grinolds2014subnanometre,doherty2014electronic,plakhotnik2014all,kucsko2013nanometre}. To fully harness the NV's sensing capabilities, providing spatially homogeneous, intense MW fields which are stable over long measurement times within well-defined volumes remains an on-going challenge. In this work, we investigate planar, micro-fabricated MW antennas to reliably deliver MW radiation to NV sensors. 

NV centers are one of the promising candidates for many quantum technological applications due to their feasible spin initialization, read-out and manipulation using laser light (mostly around 532 nm) and MW fields \cite{doherty2013nitrogen}. An intersystem crossing in the NV center's energy level scheme allows for spin state initialization by optical pumping and optical readout via spin dependent fluorescence \cite{goldman2015state,goldman2015phonon}. The effect is known as optically detected magnetic resonance (ODMR, for details refer to \cite{Barry2020,bernardi2017nanoscale}). ODMR using continuous MW is the most basic sensing mode using NVs (DC magnetometry). AC magnetometry employs pulsed MW sequences e.g. spin-echo \cite{hahn1950spin} or more advanced dynamical decoupling protocols \cite{Carr1954,gullion1990new,viola1998dynamical,cywinski2008enhance,slichter2013principles,degen2017quantum}.

Sensing with NVs can be typically divided into two approaches, either exploiting the atomic scale size of a single NV center to enable high resolution magnetometry or sacrificing the latter in favor of highly sensitive magnetometry with enhanced signal to noise ratio and sensitivity using an ensemble of NV centers. Sensing using ensembles requires highly homogeneous MW fields as any field inhomogeneity over the ensemble constitutes a dephasing and consequently reduces sensitivity \cite{turner2020magnetic,mizuno2020simultaneous,horsley2018microwave,glenn2017micrometer,simpson2016magneto,clevenson2015broadband,nowodzinski2015nitrogen,chipaux2015magnetic,le2013optical,steinert2013magnetic,pham2011magnetic,steinert2010high,maertz2010vector}. Also for single NV-based schemes, spatially homogeneous fields are relevant especially, when scanning NV centers are used to image samples \cite{ariyaratne2018nanoscale,bernardi2017nanoscale,thiel2016quantitative,pelliccione2016scanned,appel2015nanoscale,tetienne2014nanoscale,grinolds2013nanoscale,maletinsky2012robust,rondin2012nanoscale}.

For AC magnetometry, the sensitivity depends on the coherence time $T_2$ which can be enhanced by employing dynamical decoupling protocols. In such multipulse protocols, short, intense MW pulses are of interest to perform full pulse sequences within the potentially low coherence and dephasing times of NV centers. This demand for homogeneous, stable and intense MW fields motivated our work for an optimized MW antenna design for NV sensing. 
\begin{figure}[ht]
\centering
\includegraphics[width = 0.9 \textwidth]{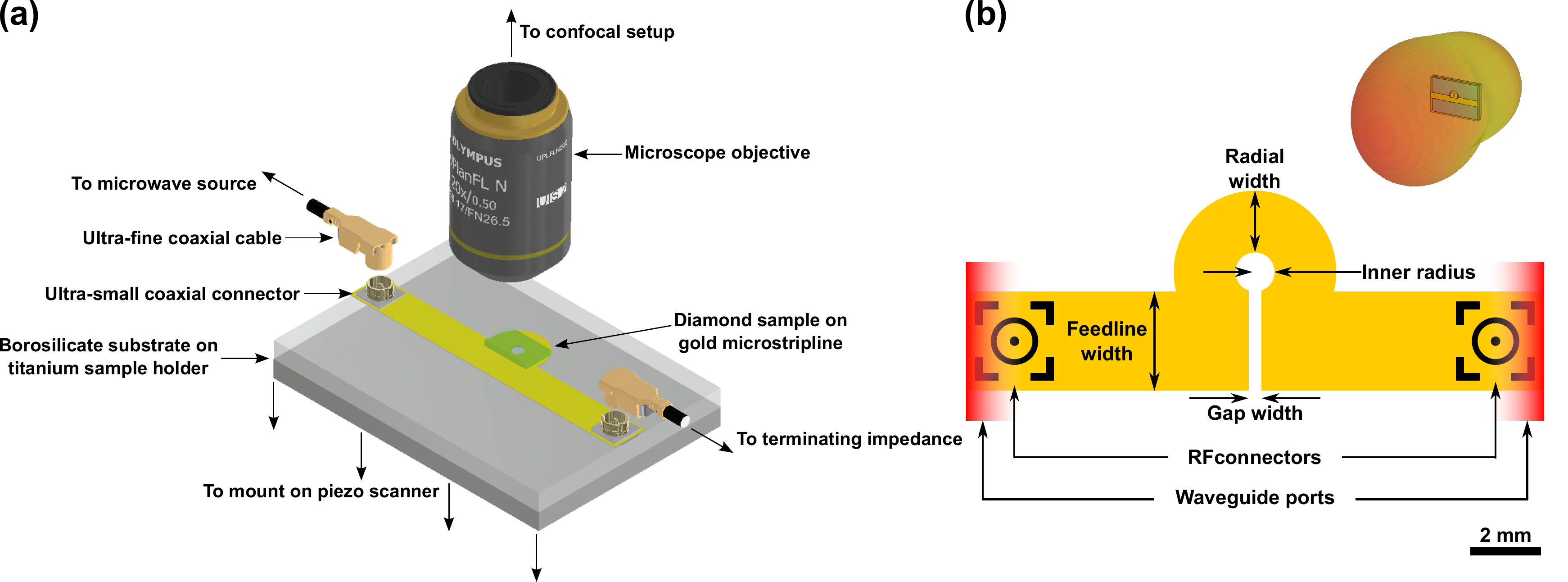}
\caption{(\textbf{a}) Schematic representation of the $\Omega$-shaped antenna and it's implementation into a confocal microscope utilizing piezo scanner (Coaxial connector design by avitek, coaxial cable design by Steven Minichiello, and objective design by thorfynn. Source: https://grabcad.com, accessed on 07.06.2021). (\textbf{b}) Schematic representation of the geometry of the $\Omega$-shaped antenna design (note that the scale bar is approximate and the sketch does not give the exact geometry of the simulated antenna but illustrates the design in general). Microwave (MW) modes enter and leave the calculation domain via the waveguide ports indicated in red. Inset: Simulated MW radiation pattern. Here, a strong directivity towards the sample is visible.}
\label{fig:Geometry}
\end{figure}

In this study, we investigate MW antenna designs compatible with confocal and atomic force microscopy (AFM) based sensing applications with NV centers [\textbf{Fig.} \ref{fig:Geometry}(\textbf{a})] \cite{ariyaratne2018nanoscale,thiel2016quantitative,pelliccione2016scanned,appel2015nanoscale,tetienne2014nanoscale,grinolds2013nanoscale,maletinsky2012robust,rondin2012nanoscale,bernardi2017nanoscale}. Commonly used antenna concepts show design related limitations with respect to these applications: wire antennas suffer from inhomogeneity in the radiated fields requiring precise positioning \cite{li2010design}, while stripline antennas tend to be large in size and overload piezo scanners in confocal microscopes when placed on top \cite{rudnicki2013microwave,mrozek2015circularly,sasaki2016broadband,qin2018near}. Coil antennas only radiate MW of low amplitude \cite{chen2018large,dong2018fiber,soshenko2018microwave} and resonators are limited in bandwidth \cite{alegre2007polarization,bayat2014efficient,herrmann2016polarization,zhang2016microwave,yang2019microstrip,mariani2020system,yaroshenko2020circularly}.

We aim for planar antenna designs that can be efficiently incorporated into sensing setups without significant hardware modification [\textbf{Fig.} \ref{fig:Geometry}(\textbf{a})]. To investigate larger microstructures as e.g. biological samples with characteristic dimensions of the order of several hundreds of $\mu$m, we aim for macroscopically large and homogeneously radiated MW fields of high amplitude. The stripline-based design we use offers risk-free sample handling by mounting the sample on top of the antenna, implementation without blocking optical access, along with the advantage of established fabrication technology. By using transparent substrates for the antenna, we provide an antenna design that also allows observation in inverted geometries where the optical path crosses the antenna or the antenna is attached to the microscope objective  \cite{horowitz2012electron,simpson2016magneto}. We use a design based on an $\Omega$-shape  [\textbf{Fig.} \ref{fig:Geometry}(\textbf{b})]. This layout inherently leads to the radiation of a homogeneous MW field within the aperture towards the designated sample volume [see inset \textbf{Fig.} \ref{fig:Geometry}(\textbf{b})].

Our manuscript is structured as follows:  \textbf{Sec.} \ref{Sec3} summarizes the process flow, relevant materials and parameters for our MW antenna fabrication. \textbf{Sec.} \ref{Sec4} summarizes the methodology, assumptions and the numerical simulation approach to find an optimized antenna design. \textbf{Sec.} \ref{Sec5} briefly describes the experimental setup and methods to characterize and test the antennas. \textbf{Sec.} \ref{Sec6} discusses our experimental findings on the MW antenna performance.

\section{Microfabrication Methods}
\label{Sec3}
\begin{figure}[ht]
\centering
\includegraphics[width = 0.9 \textwidth]{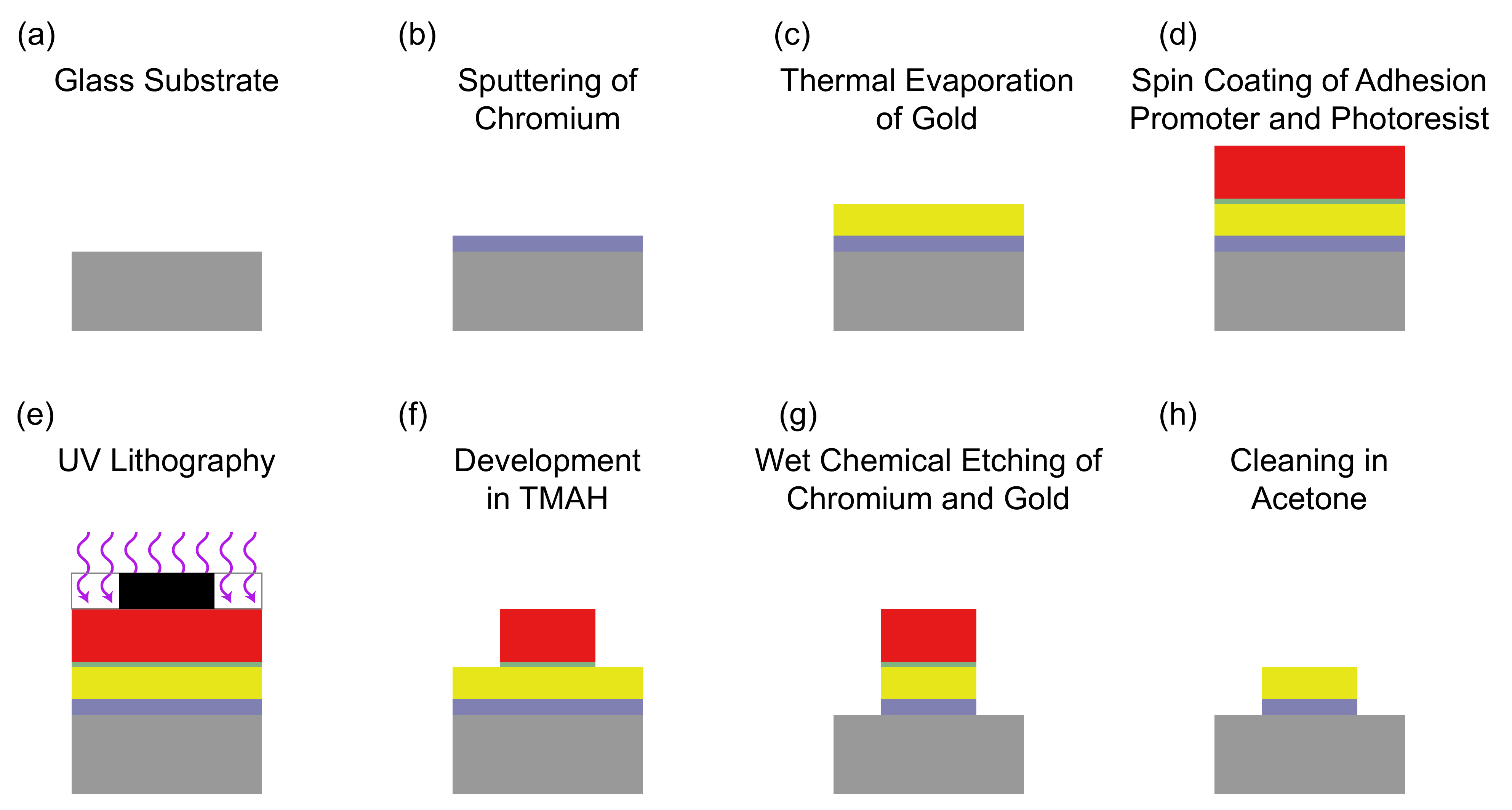}
\caption{Process flow for Omega antennas. The different color correspond to: Grey – Glass
Substrate, Blue – Chromium, Yellow – Gold, Green – Adhesion promoter, Red –
Photoresist, Black – Photomask, Violet – UV-Illumination.}
\label{fig:Workflow}
\end{figure}

We fabricated the antennas on borosilicate glass substrates [Relative permittivity $\epsilon_r$ = 4.82 , Substrate thickness $s_z$ =  1 mm, \textbf{Fig.} \ref{fig:Workflow}(\textbf{a})]. To prepare thin film deposition, we cleaned the substrate in acetone followed by isopropanol using an ultrasonic bath. We removed absorbed water by heating the substrates on a heatplate (120 $^\circ$C, 10 min). We then used either sputtering or thermal evaporation to deposit an adhesion layer (20 nm chromium) [\textbf{Fig.} \ref{fig:Workflow}(\textbf{b})] followed by a 100 nm gold layer [\textbf{Fig.} \ref{fig:Workflow}(\textbf{c})]. To fabricate the antenna structure, we used spin coating (6000 rpm, 1 min) followed by baking (120 $^\circ$C, 2 min) to deposit an adhesion promoting sub-monolayer of TI Prime (MicroChemicals, Ulm, Germany). We then spin coated the photoresist AZ1518 (6000 rpm, 1 min, Merck, Darmstadt, Germany) [\textbf{Fig.} \ref{fig:Workflow}(\textbf{d})]. After a prebake (100 $^\circ$C, 50 s), we applied contact UV lithography using a laser written binary intensity amplitude chromium photomask [\textbf{Fig.} \ref{fig:Workflow}(\textbf{e})]. Lithography parameters were chosen according to resist manufacturer specifications resulting in an area dose of 33.6 $\frac{\text{mJ}}{\text{cm}^2}$. We developed the resist mask by stirring the exposed substrates in TMAH (Acros Organics, Fair Lawn, USA) 2.5 \% in water [\textbf{Fig.} \ref{fig:Workflow}(\textbf{f})]. After rinsing it in monodistilled water, we performed a postbake (120 $^\circ$C, 50 s). To form the antenna structure using wet chemical etching: First, we stirred the substrate in gold etchant (Sigma-Aldrich, St. Louis, USA) followed by rinsing it in chromium etchant [Sigma-Aldrich, \textbf{Fig.} \ref{fig:Workflow}(\textbf{g})]. To remove residual photoresist and potential contamination, we stirred the fabricated MW antenna in acetone and isopropanol [\textbf{Fig.} \ref{fig:Workflow}(\textbf{h})]. This allowed electrical contacting by gluing SMT ultra-miniature coaxial connectors [U.FL-R-SMT(01), Hirose Electronic, Tokyo, Japan] using electrically conductive epoxy adhesives (EPO-TEK H20E, Epoxy Technology, Billerica, USA).

\section{Numerical Simulation and Optimization}
\label{Sec4}
For the numerical optimization, we use the time domain solver of the commercial software CST Microwave Studio (Dassault Systèmes, Vélizy-Villacoublay, France) which uses the finite integral technique (FIT) to solve the Maxwell equations \cite{weiland1977discretization}. We define waveguide ports at the ends of the microstripline, where waveguide modes can leave or enter the domain of interest. During each simulation, the in total delivered MW power at the excited port equals 500 mW. We typically run the simulation in the frequency range between 2 GHz and 4 GHz. To enhance far field calculation accuracy, our simulation includes additional space between the structure and the simulated space boundaries ($\frac{\lambda_\text{MW}}{4} \approx $ 26 mm). The boundaries fulfill the perfectly matched layer (PML) condition assuring that incident waves are being absorbed with negligible back reflections. The applied mesh exhibits variable mesh cell sizes and is locally refined by introducing additional cells within the space around critical structural features like corners, edges and material interfaces. This approach allows for highly precise simulations while keeping computational time reasonable. Due to the varying structure geometries, the simulated space is typically meshed with approximately 180 $\times$ 190 $\times$ 140 cells resulting in 4.6 million cells involving a total volume of roughly 68 mm $\times$ 63 mm $\times$ 54 mm.

\begin{figure}[ht]
\centering
\includegraphics[width = 1 \textwidth]{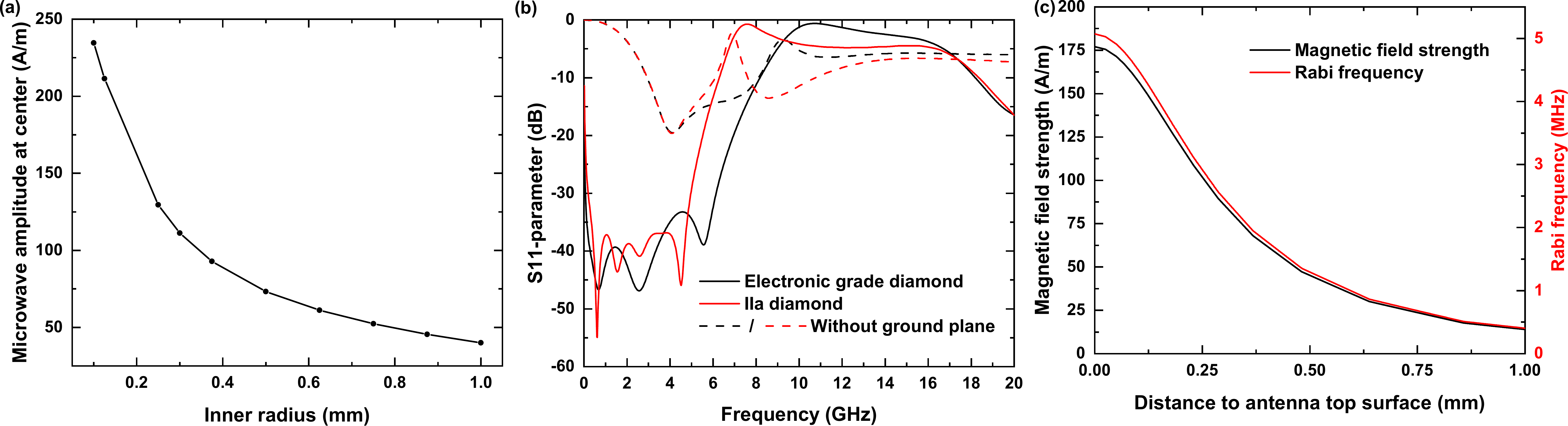}
\caption{(\textbf{a}) Achievable MW amplitude for different inner radius values $r_w$. Points show from different simulation runs determined data. (\textbf{b}) Calculation of the frequency-dependent $S_{11}$ parameter (back reflection coefficient) with ground plane for 50 $\mu$m thick diamond (solid black) and 300 $\mu$m thick diamond (solid red) and without ground plane (dashed lines) respectively. (\textbf{c}) MW amplitude and Rabi frequeny dependence on the distance to the antenna top surface. Rabi frequencies $\Omega_\text{R}$ are determined for a resonantly driven transition of the subset of nitrogen vacancy (NV) centers aligned in $\langle 1 1 1 \rangle$ crystallographic direction. Input MW power equals 1 W.}
\label{fig:InnerRadiusS11}
\end{figure}

The linear and radial parts of the antenna form capacitive and inductive components respectively. Altogether, they determine the antenna's resonance. To harness this antenna design for NV-based sensing, we optimize the linear and radial geometry to match antenna resonance and NV spin transition frequency. Consequently, the main parameters to be optimized are the gap width $g_w$ and the radial width $r_w$ of the $\Omega$-antenna [see \textbf{Fig.} \ref{fig:Geometry}(\textbf{b})]. In addition, optimizing the feedline width $f_w$ also yields a slight improvement. We also considered the thickness of the gold layer $t_\text{Au}$ forming the conductor. Our results indicate a negligible influence on resonance frequency and field distribution for films thinner than 1 $\mu$m. Furthermore, we do not observe significant improvement capabilities by increasing the layer thickness further. All simulated gold layers are thinner than the theoretically expected skin depth for gold of 1.4 $\mu$m at 2.87 GHz leading to an approximately homogeneous current density throughout the gold layer. We assume the contribution of the chromium adhesion layer to the antenna properties to be negligible. We thus do not vary the thickness of this layer and set it to 20 nm.

To achieve high MW field within the aperture, the inner radius of the aperture should be minimized contradicting the aim for a macroscopically large radiated area. For all optimization steps, we evaluate the magnetic field amplitude at the center of the antenna's aperture at a distance of 10 nm below the diamond sample's top surface, which corresponds to the approximate depth of shallow NV centers. For the diamond sample, we consider a (100)-oriented diamond (2.5 mm $\times$ 1.5 mm $\times$ 50 $\mu$m). Typically, for experiments using single NV centers, high purity electronic grade, chemical vapor deposited diamonds (substitutional nitrogen content $<$ 5 ppb, manufacturer: Element Six, Didcot, United Kingdom) are used. For testing the antenna, we use less pure diamonds (substitutional nitrogen content $<$ 1 ppm, termed IIa diamonds) which are 300 $\mu$m thick and contain a dense ensemble of NV centers, which we also consider for better comparison with our experimental results. Note that for the simulations, the dielectric constant used for electronic grade and IIa diamonds is identical as it is not influenced by the nitrogen content. By varying the inner radius of the aperture, we find a trade-off between the achievable MW amplitude and the radiated area size as shown in \textbf{Fig.} \ref{fig:InnerRadiusS11}(\textbf{a}). We choose the inner radius to be 300 $\mu$m, which is smaller than the diamond under consideration, but allows to investigate reasonably large sample areas. Note that only for an inner radius value of 300 $\mu$m the design was fully optimized. For all other values, we expect the amplitude can be further enhanced subjecting to the repeated optimization of the remaining parameters. Otherwise, these calculations assumed the same parameter values we determined during the optimization for an inner radius of 300 $\mu$m.

We set the lateral substrate size to 16 mm $\times$ 11 mm. This substrate allows for straightforward mounting to typical piezo-scanner in confocal microscope setups. The simulation also includes the sample holder which we assume to be a 24 mm $\times$ 15 mm titanium plate as compliant with the piezo scanner systems (attocube systems, Haar, Germany) used in our setup.

For the optimization, we apply the software's internal goal function by defining quantities and frequency ranges in which they are evaluated, type of goal (minimization/maximization) and weight factors. We consider different aspects relevant for the antenna performance. The predominant limitation in maximizing the radiated field amplitude is given by frequency mismatch leading to significant back reflections of the input MW signal characterized by the high frequency circuitry component's S$_{11}$-parameter. In accordance with other reported work, we minimize the S$_{11}$-parameter of the antenna and consequently the back reflection of the input signal at 2.87 GHz \cite{sasaki2016broadband,qin2018near,chen2018large,yang2019microstrip,bayat2014efficient,herrmann2016polarization,zhang2016microwave,yaroshenko2020circularly}. Optimizing the $\Omega$-antenna for NV sensing involving variable, high magnetic bias fields, as well as for obtaining complete ODMR spectra, requires us to maximize the antenna bandwidth. To account for that, we extend the minimization of S$_{11}$ to the frequency range between 2.77 GHz and 2.97 GHz. Overall, we perform two evaluations of S$_{11}$, once individually at 2.87 GHz, and again in the range between 2.77 GHz and 2.97 GHz. Both the evaluations were given equal weight allowing us to maximize the bandwidth and simultaneously favor a centering of the resonance at 2.87 GHz. Furthermore, our optimization goal function aims to maximize the magnetic field strength. Weight factor is chosen accordingly to allow for equal contributions between the S$_{11}$-parameter and the aimed microwave amplitude.

Under these assumptions, we determine the following set of optimized geometry parameters: gap width $g_w = 7$ $\mu$m , radial width $r_w = 1.151 $ mm, feedline width $f_w = 1.851 $ mm. This antenna structure exhibits multiple resonances at 0.7 GHz, 2.6 GHz and 5.5 GHz. For the most relevant resonance at 2.5 GHz, we find a value of -47 dB as illustrated by the $S_{11}$-parameter curve in \textbf{Fig.} \ref{fig:InnerRadiusS11}(\textbf{b}). For the optimized case of 50 $\mu$m thick electronic grade diamonds, the bandwidth for this structure amounts to 8.2 GHz enabling reliable NV spin manipulation involving bias fields up to 190 mT. For 300 $\mu$m thick IIa diamonds, our results indicate a lowering of the bandwidth down to 6.3 GHz consequently limiting feasible bias fields to 120 mT. Our titanium sample holder acts as a ground plane and significantly decreases back reflections. To estimate the $\Omega$-antenna's robustness against geometry deviations arising from UV lithography tolerance errors, we repeat the simulation and adjust either the gap width by $\pm$ 1 $\mu$m or the radial width by $\pm$ 10 $\mu$m and compare the respective $S_{11}$-parameter curves. We find slight resonance shifts of approximately -50 MHz / $\mu$m for increasing gap width and -20 MHz / $\mu$m for increasing radial width. Together with its wideband resonance property, this design is suited for versatile applications in the field of NV based sensing \cite{turner2020magnetic,horsley2018microwave,simpson2016magneto,clevenson2015broadband,nowodzinski2015nitrogen,chipaux2015magnetic,le2013optical,steinert2013magnetic,pham2011magnetic,steinert2010high,maertz2010vector,ariyaratne2018nanoscale,thiel2016quantitative,pelliccione2016scanned,tetienne2014nanoscale,grinolds2013nanoscale,maletinsky2012robust,rondin2012nanoscale}.

\begin{figure}[ht]
\centering
\includegraphics[width = 1 \textwidth]{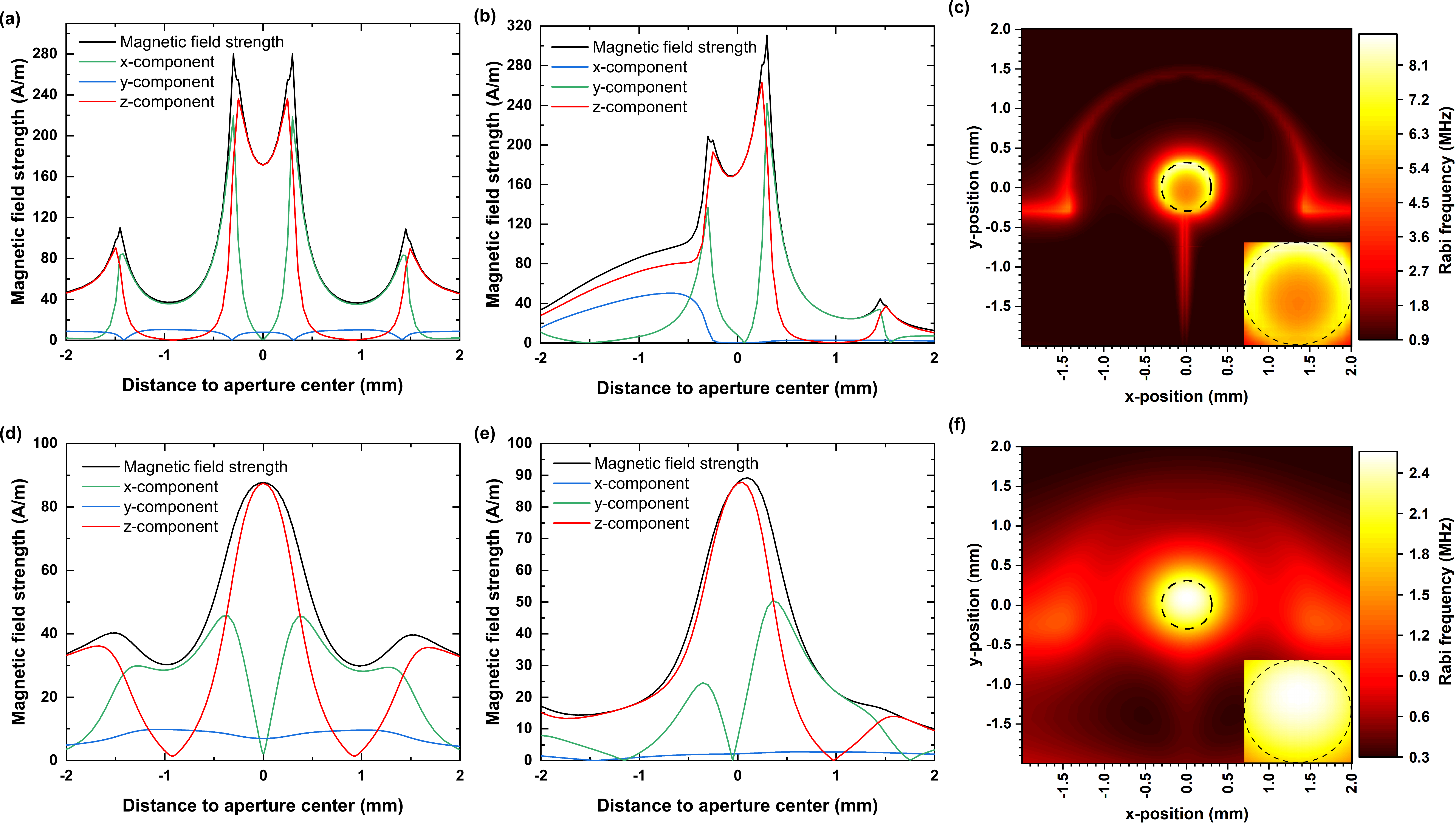}
\caption{Field simulation results for 50 $\mu$m thick electronic grade diamond (\textbf{a}), (\textbf{b}) and (\textbf{c}) and for 300 $\mu$m thick IIa diamond (\textbf{d}), (\textbf{e}) and (\textbf{f}) respectively. All field data were taken 10 nm below the diamond sample's top surface and correspond to a MW input power of 1 W. (\textbf{a}), (\textbf{b}), (\textbf{d}) and (\textbf{e}) Field distribution of the MW in x and y-direction. (\textbf{c}) and (\textbf{f}) 2D maps of theoretically expected Rabi frequencies $\Omega_\text{R}$ for a resonantly driven transition of the subset of nitrogen vacancy (NV) centers aligned in $\langle 1 1 1 \rangle$ crystallographic direction. The dashed circles mark the circumference of the aperture. Insets: Detail views of the aperture areas.}
\label{fig:FieldDistributionRabiMap.pdf}
\end{figure}

With increasing distance to the antenna's top surface we find a decrease in MW field amplitude [see \textbf{Fig.} \ref{fig:InnerRadiusS11}(\textbf{c})]. For NV-Centers very close to antenna's top surface the achievable Rabi frequencies at the aperture center reach values up to 5.1 MHz. For coherent spin manipulation of shallow NV centers we find in case of 50 $\mu$m electronic grade diamonds Rabi frequencies of 4.9 MHz and 2.5 MHz for 300 $\mu$m thick IIa diamonds respectively. Due to an approximate radial symmetry of the $\Omega$-antenna, the established MW field distribution shows a radial distance dependence from the aperture center in the x-y plane [see \textbf{Fig.} \ref{fig:FieldDistributionRabiMap.pdf}(\textbf{a}) and \ref{fig:FieldDistributionRabiMap.pdf}(\textbf{b})]. For an electronic grade diamond with 50 $\mu$m thickness, the typically achieved value of the magnetic field amplitude exceed 170 $\frac{\text{A}}{\text{m}}$. Towards the aperture circumference, we observe a parabolic increase up to 280 $\frac{\text{A}}{\text{m}}$. Within an area with a radius of 260 $\mu$m, the total MW field is mainly given by its perpendicular z-component as shown in \textbf{Fig.} \ref{fig:FieldDistributionRabiMap.pdf}(\textbf{a}) and \ref{fig:FieldDistributionRabiMap.pdf}(\textbf{b}). Beyond the circumference, there is a further increase in the amplitude due to a strong increase in the lateral field components, in contrast the z-component drops off quickly. In the direction of the gap, the MW field amplitudes do not reach the mentioned peak values and decrease near it instead. From the field distribution, we calculate the expected Rabi frequency $\Omega_\text{R}$ considering detuning due to field components parallel to the NV axis for one subset of NV centers [see \textbf{Fig.} \ref{fig:FieldDistributionRabiMap.pdf}(\textbf{c})]. For a square area fitting into the aperture (edge length 420 $\mu$m) we determined an average value for the expected $\Omega_\text{R}$ of 5.9 MHz $\pm$ 0.5 MHz. For later comparison with our experimental data \textbf{Fig.} \ref{fig:FieldDistributionRabiMap.pdf}(\textbf{d}),(\textbf{e}) and (\textbf{f}) show the obtained simulation data in case for an IIa diamond with 300 $\mu$m thickness. In that case we determined an average value for the expected $\Omega_\text{R}$ of 2.38 MHz $\pm$ 0.06 MHz.

\section{Experimental Setup and Methods}
\label{Sec5}
To asses the performance of the microfabricated antennas, we apply them to typical methods used in NV based sensing (see Ref. \cite{Barry2020,bernardi2017nanoscale}). We mount the antenna on a sample holder using a thin layer of Crystal Bond 509 (Structure Probe Inc, West Chester, USA). As sample, we used a single crystalline, (100)-oriented IIa, chemical vapor deposited diamond (Element Six, 3 mm $\times$ 3 mm $\times$ 300 $\mu$m, substitutional nitrogen content $<$ 1 ppm). We cleaned the diamond in a tri-acid mixture (H$_2$SO$_4$ 96 \%, HClO$_4$ 70 \%, HNO$_3$ 65 \%) at 500 $^\circ$C for 1 h. The sample contains a homogeneous ensemble of native NV centers as confirmed by multiple photoluminescence confocal scans. 

We use a lab-built confocal microscope setup (numerical aperture NA = 0.8, excitation wavelength $\lambda_{Laser}$ = 532 nm). Confocal filtering of the NV ensemble's fluorescence was realized by using a single mode optical fiber. To measure the fluorescence, we use a single photon detector (SPCM-AQRH-14, quantum efficiency $\approx$ 68\%, Excelitas Technologies, Waltham, USA) and a data acquisition card for acquiring, and logging the signal (PCIe-6323, National Instruments, Austin, USA).

We connect the $\Omega$-antennas to a MW source (SG 384, Stanford Research Systems, Sunnyvale, USA) with an amplifier (ZASWA-2-50DR+, typ. +30 dB, Mini Circuits, Brooklyn, USA). We apply a static bias field by positioning a neodymium (NdFeB) permanent magnet using a three axis linear stage. We align the bias field  with one of the $\langle 1 1 1 \rangle$ crystal directions within an error of $\approx 15 \, ^\circ$.

We realize Rabi oscillations (see Ref. \cite{Jelezko2004}), where we determine the achievable Rabi frequency $\Omega_\text{R}$ in the NV spin system with the MW power deployed by the antenna. The $\Omega_\text{R}$ is the prime measure for antenna performance because it depends linearly on the perpendicular component of the MW's amplitude relative to the NV axis. These measurements were done for different positions to map the spatial homogeneity of the Rabi frequency. We also demonstrate continuous wave and pulsed optically detected magnetic resonance (ODMR) measurements (see Ref. \cite{Gruber2012}) as well as dynamical decoupling protocols (see Ref. \cite{MacQuarrie2015}) showing the suitability of the MW antenna design for nanoscale NV based sensing applications.

To generate the control MW pulses used in dynamical decoupling experiments, we use the in-built digital I/Q mixer of the MW signal generator in combination with external quadrature signal sources (Pulse Streamer 8/2, Swabian Instruments, Stuttgart, Germany). With this approach, we control the pulse duration, shape and polarity of the MW signal. For dynamical decoupling experiments, we use a stronger amplifier (ZHL-16W-43-S+, typ. +45 dB, Mini-Circuits) and a 520 nm diode laser (DL nSec, PE 520, Swabian Instruments). 

\section{Antenna Performance}
\label{Sec6}
Comparison - Sputtering \& Thermal Evaporation: We investigated the influence of the physical vapor deposition (PVD) method used to deposit the electrically conducting gold layer. For this purpose, we characterized two geometrically identical MW antenna batches, one with a sputtered gold layer the other with a thermally evaporated gold layer. We note that the sputtered gold layer shows superior adhesion compared to the thermally evaporated one.

\begin{figure}[ht]
\centering
\includegraphics[width = 0.9 \textwidth]{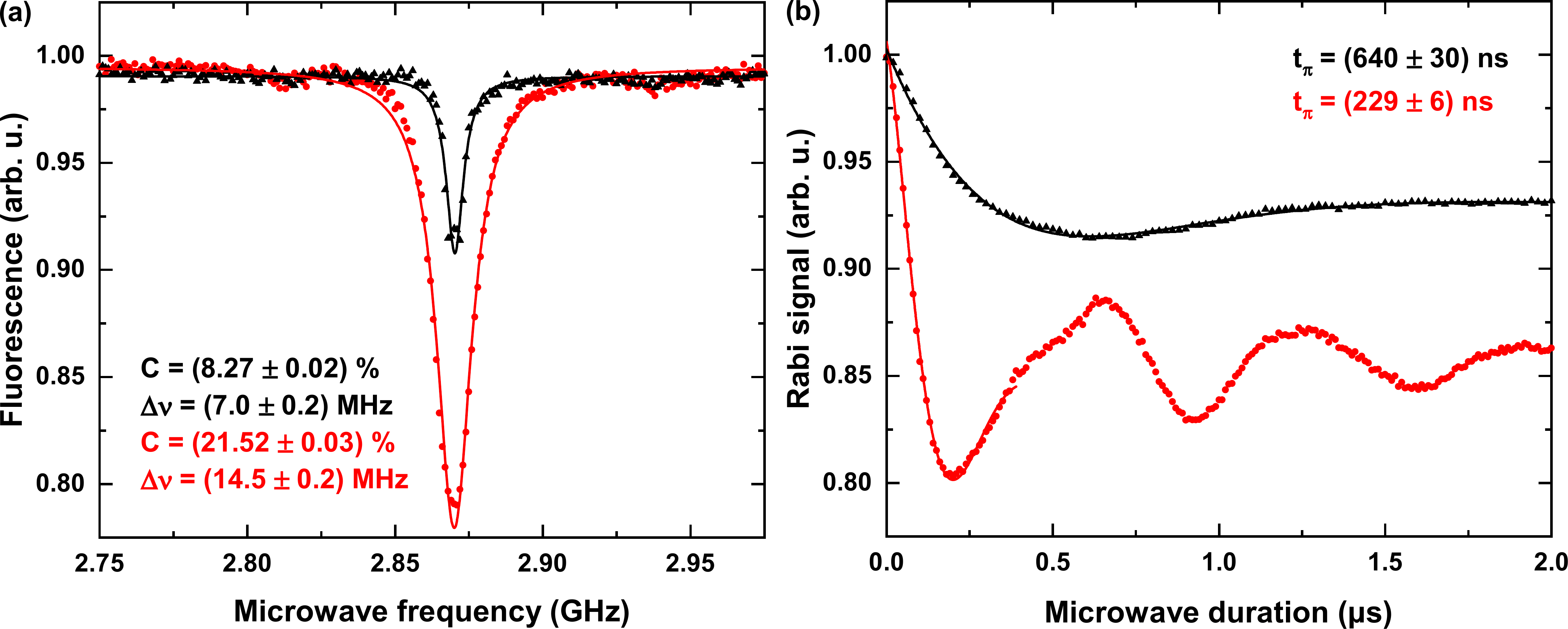}
\caption{Comparison between sputtered and evaporated layers. The evaporated antennas show a better performance in terms of radiated MW amplitude as shown by the increased optically detected magnetic resonance (ODMR) contrast C [(\textbf{a}), 8 \% vs 22 \%], and Rabi frequency $\Omega_\text{R}$ [(\textbf{b}), 800 kHz vs 2.2 MHz, $t_{\pi}$= $\pi$-pulse duration]. Note that the ODMR resonance of the evaporated antenna is strongly power broadened ($\Delta \nu$ = FWHM). Points indicate the experimentally obtained data and solid curve denotes fits used to obtain the mentioned parameters.}
\label{fig:SputterVSEvaporatePanel}
\end{figure}

We performed ODMR and Rabi oscillations without an external magnetic field. The optical excitation power was set to $P_\text{Opt}$ = 700 $\mu$W and the MW power before amplification (+45 dB) was $P_\text{MW}$ =  -15 dBm. We find that antennas with thermally evaporated gold layer outperform antennas with sputtered layer as shown in \textbf{Fig.} \ref{fig:SputterVSEvaporatePanel}(\textbf{a}). On average, we determined ODMR contrasts of 10 \% $\pm$ 3 \% and $\Omega_\text{R}$ of 0.86 MHz $\pm$ 0.06 MHz for sputtered antennas and 21 \% $\pm$ 2 \% contrast and $\Omega_\text{R}$ of 2.4 MHz $\pm$ 0.5 MHz for thermally evaporated ones. The thermally evaporated antennas achieve comparably high contrasts in comparison to highly optimized ODMR experiments ($ \approx $ 30 \% according to \cite{mizuno2021electron,osterkamp2019engineering}) and significant power broadening of the ODMR resonances indicates high MW power driving the NV centers as displayed in \textbf{Fig.} \ref{fig:SputterVSEvaporatePanel}(\textbf{b}). We attribute this enhancement to a lower power loss within the gold layer due to a decreased amount of incorporated impurities and crystal defects created during the sputtering process. Taking into account this finding, we restrict further characterization to thermally evaporated antennas.

\begin{figure}[ht]
\centering
\includegraphics[width = 0.9 \textwidth]{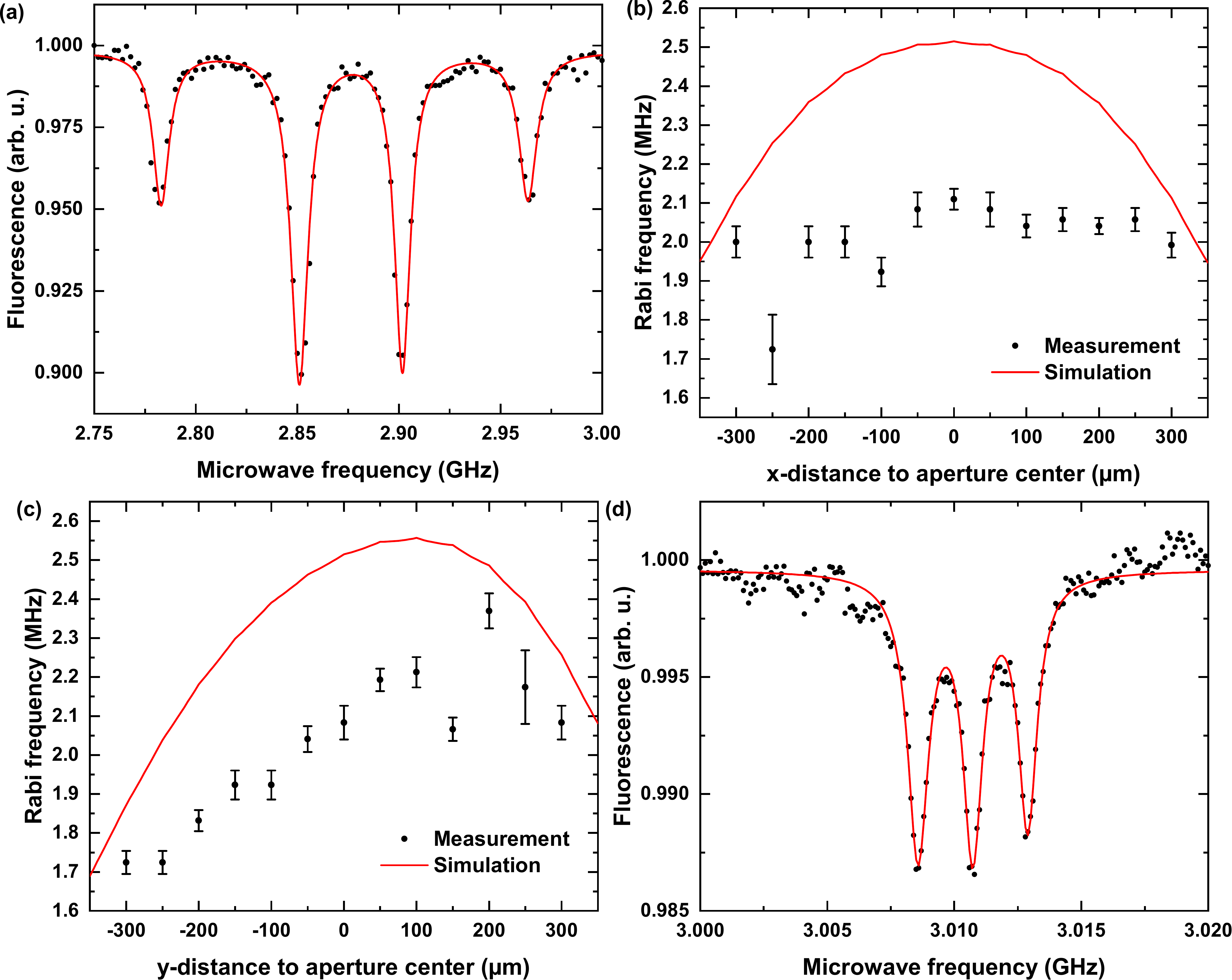}
\caption{(\textbf{a}) Continuous ODMR measurement of NV ensemble with an externally applied magnetic field. For the outermost resonance pair, a splitting of 181 MHz $\pm$ 1 MHz with high contrast (4.67 \% $\pm$ 0.06 \% at -15 dBm source MW power) is obtained. For the corresponding subset of NV centers, we determined a projection of the magnetic field on the NV axis of 6.45 mT $\pm$ 0.05 mT.  (\textbf{b})/(\textbf{c}) Characterization of the homogeneity of the radiated MW field amplitude perpendicular and parallel to the gap. While the amplitude parallel to the gap slightly depends on the distance to the gap as expected, the amplitude perpendicular to the gap remains unaffected showing that the full area of the aperture is reliably usable for spin manipulation protocols. (\textbf{d}) Pulsed ODMR of one of the resonances showing the hyperfine transitions due to $^{14}N$ nuclear spin coupling. The black points indicate the experimentally obtained data, solid black lines indicate the error bars and the solid red lines indicate either fits or simulated data.}
\label{fig:ExperimentalResults}
\end{figure}

Spatial Homogeneity: To evaluate the spatial homogeneity of the radiated MW field, we performed Rabi oscillation measurements ($P_\text{Opt}$ = 700 $\mu$W , $P_\text{MW}$ =  -15 dBm) along multiple designated line-cuts in x and y direction respectively (perpendicular/parallel to the gap). For these measurements, we apply an external magnetic field mostly aligned with one of the $\langle 1 1 1 \rangle$ crystallographic directions. Thus, leading to a significant Zeeman splitting for that subset of NV centers \cite{bernardi2017nanoscale}. In this situation, ODMR contrast in a non-aligned ensemble intrinsically lowers. For this resonance pair, we observe 4.67 \% $\pm$ 0.06 \% contrast which is typical for NV ensembles involving not perfectly aligned bias fields leading to spin-mixing as well as charge state instabilities \cite{mizuno2021electron,osterkamp2019engineering}. Rabi oscillations were driven typically at a transition frequency of 2.783 GHz like in \textbf{Fig.} \ref{fig:ExperimentalResults}(\textbf{a}) corresponding to a bias field strength of $B_\text{NV}$ = 6.45 mT $\pm$ 0.05 mT. \textbf{Fig.} \ref{fig:ExperimentalResults}(\textbf{b}) and (\textbf{c}) summarize the measurements of the spatial homogeneity of the $\Omega_\text{R}$. Line-cuts perpendicular to the gap reveal constant $\Omega_\text{R}$ within experimental errors. As expected from the simulation results, we observe a slight increase in $\Omega_\text{R}$ with increasing distance from the gap opening. We determine an average value of 2.1 MHz $\pm$ 0.1 MHz over all measurement points within 200 $\mu$m distance to the aperture center. Homogeneity and range are in good agreement with our simulation results [compare \textbf{Fig.} \ref{fig:FieldDistributionRabiMap.pdf}(\textbf{f})]. We attribute the in average slightly lowered Rabi frequency $\Omega_R$ to unconsidered losses of our MW circuitry and the potential impedance mismatch of the electrically conducting gluing point for the coaxial connectors. The radiated MW fields homogeneity on the scale of the aperture area enables higher coherence times for NV ensembles leading to enhanced sensitivities. Furthermore, simultaneous MW irradiation of high amplitude within a macroscopic area provides position independent high contrast and therefore improved signal-to-noise ratio making this antenna type promising for wide-field \cite{turner2020magnetic,mizuno2020simultaneous,horsley2018microwave,glenn2017micrometer,simpson2016magneto,clevenson2015broadband,nowodzinski2015nitrogen,chipaux2015magnetic,le2013optical,steinert2013magnetic,pham2011magnetic,steinert2010high,maertz2010vector} as well as for scanning NV based sensing applications \cite{ariyaratne2018nanoscale,bernardi2017nanoscale,thiel2016quantitative,pelliccione2016scanned,appel2015nanoscale,tetienne2014nanoscale,grinolds2013nanoscale,maletinsky2012robust,rondin2012nanoscale}.

Suppression of power broadening - Pulsed ODMR: NV related applications often require to suppress the power broadening e.g. when hyperfine transitions of the NV center need to be addressed. Pulsed ODMR schemes instead of continuous driving can suppress power broadening \cite{dreau2011avoiding}. Here, we show that our antenna also allows for reliable pulsed ODMR measurements. We perform pulsed ODMR using NV centers close to the  aperture center. Optical spin initialization and read out at 532 nm was realized with 1 $\mu$s long laser pulses. Resonant spin manipulation involved 2 $\mu$s MW pulses at -30 dBm source MW power amplified by factor of +45 dB. The pulsed ODMR measurement at $B_\text{NV}$ = 5 mT in \textbf{Fig.} \ref{fig:ExperimentalResults}(\textbf{d}) reveals the hyperfine interaction of the NV center electronic spin with the $^{14}$N nucleus. We thus confirm usability of our antennas at low power for efficient, pulsed spin manipulation. 

\begin{figure}[ht]
\centering
\includegraphics[width = 0.9 \textwidth]{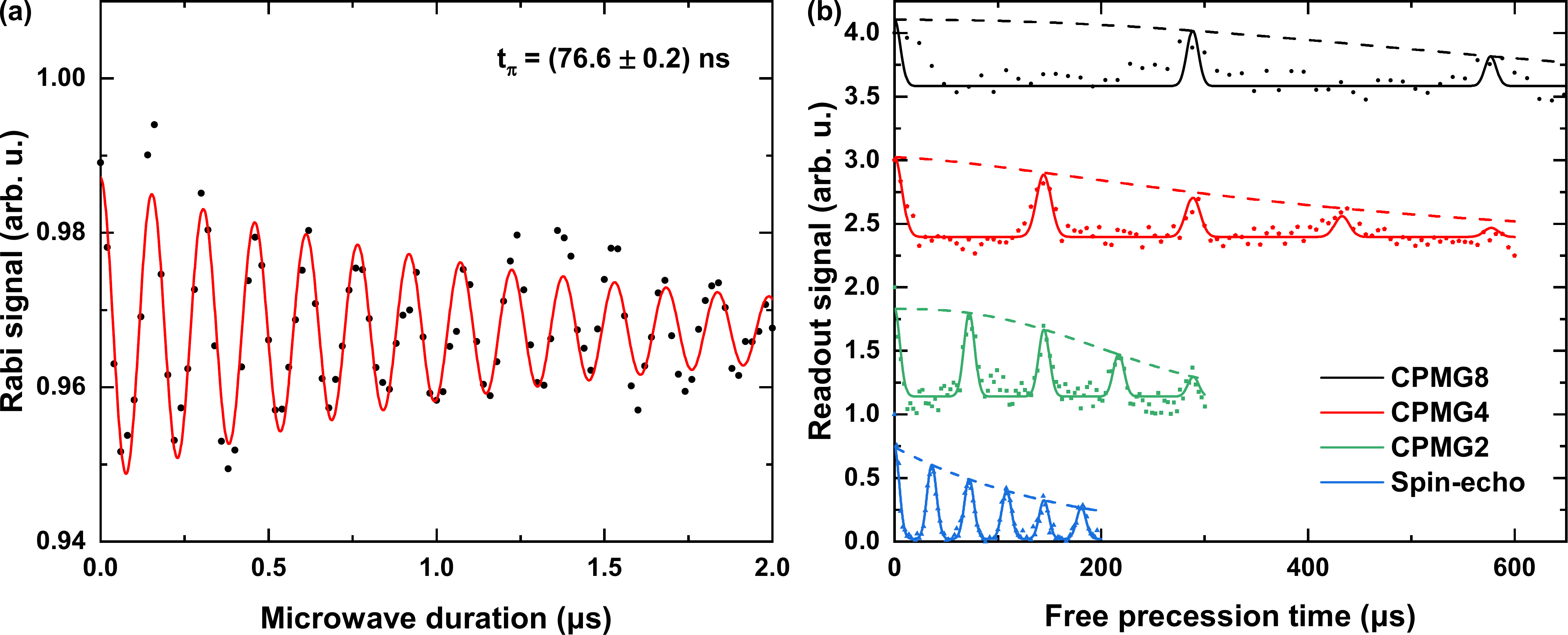}
    \caption{(\textbf{a}) Typical Rabi oscillation measurement which is primarily used to obtain relevant pulse parameter to be used in multi-pulse sensing schemes. Measurement performed at $\Omega_R \approx$  6.5 MHz. (\textbf{b}) Spin-echo and Carr Purcell Meiboom Gill (CPMG)-n  measurements with the NV ensemble performed with the pulse parameters obtained from (\textbf{a}). The plots have been fit and scaled along the y-axis for comparison. The points show the experimentally obtained data, solid curves denote the fit, and the dashed lines indicate the fit envelopes. The spin-echo measurements reveals a $T_2 \approx$ 167 $\mu$s. Applying CPMG-8 pulse sequences leads to almost a fourfold improvement in the decoherence time of the spin state. From the fit, the calculated enhanced decoherence times for the CPMG-2/4/8 protocols are $\approx$ 245 $\mu$s / $\approx$ 425 $\mu$s / $\approx$ 638 $\mu$s respectively.}
    \label{fig:DynamicalDecoupling}
\end{figure}

High Power Applications: High MW powers enable to realize faster and more complex pulse sequences typically occurring in dynamical decoupling schemes \cite{degen2017quantum}. Dynamically decoupling the NV center from the environmental noise can increase the coherence times resulting in higher AC magnetic field sensitivities and an increase of the detectable frequency range of AC magnetic fields. Here, we demonstrate dynamical decoupling using spin-echo \cite{hahn1950spin} and Carr Purcell Meiboom Gill (CPMG) pulse sequences \cite{Carr1954}. We resonantly drive a transition at 3.01076 GHz corresponding to $B_\text{NV}$ = 5 mT. Note that we did not observe any significant heating or evident nonlinear effects when applying high power MW pulses. We observed $\Omega_\text{R}$ in the range of 6 MHz to 10 MHz, depending on the control power amplification [\textbf{Fig.} \ref{fig:DynamicalDecoupling}(\textbf{a})]. With the MW power applied in \textbf{Fig.} \ref{fig:DynamicalDecoupling}(\textbf{a}), we realized spin-echo and CPMG pulse sequences with different number of $\pi$-Pulses (1, 2, 4 and 8 respectively) [\textbf{Fig.} \ref{fig:DynamicalDecoupling}(\textbf{b})]. In comparison to spin-echo which yielded $T_2$ = 167.1 $\mu$s, CPMG8 allowed to increase $T_2$ up to 638.1 $\mu$s. The antenna is evidently suitable for high power applications especially for NV based AC magnetometry. We infer that our antennas are also usable for advanced spin manipulation techniques based on quantum optimal control theory \cite{Rembold2020}.

\section{Conclusion}
\label{Sec7}
In this work, we designed and fabricated MW antenna systems for NV based sensing applications. We optimized a $\Omega$-shaped, microstripline design considering macroscopic spatial homogeneity, wide bandwidth, large amplitudes, and ease of implementation with scanning confocal/AFM setups, risk-free sample handling and straightforward fabrication. Our simulations show that the gap and the radial structure of the antenna govern its resonance properties. Key feature of the design involve an aperture with a radius of 300 $\mu$m defining the homogeneously irradiated area. The developed design achieves a high bandwidth along with desirable resonance properties within a macroscopic area with considerably high MW field amplitudes well-suited for NV based sensing applications. We established a reliable microfabrication process and tested the fabricated MW antennas for typical NV spin manipulation protocols.

We compared sputtering and thermal evaporation for the deposition of the electrically conducting gold layer and found that thermally evaporated layers significantly improve the antenna performance. By mapping the achievable Rabi frequencies within the antenna aperture, we show the macroscopic homogeneity of the radiated MW field. We test these antennas in low and high power applications by demonstrating pulsed ODMR measurements and advanced dynamical decoupling protocols. Our antennas can be straightforwardly used in optimized NV sensing approaches e.g. using optimal control pulses as well as in scanning NV magnetometry setups. Consequently, we envisage that our antenna design can add to practical NV sensing.

\section*{Acknowledgements}
We would like to acknowledge Prof. Dr. Uwe Hartmann (Saarland University, Saarbrücken, Germany) for granting access to their cleanroom facilities, especially the sputtering machine, Indujan Sivanesarajah (Saarland University, Saarbrücken, Germany) for his assistance, Günter Marchand (Saarland University, Saarbrücken, Germany) for his help in fabricating the chromium photomask, Robert Strahl (Leibniz Institute for New Materials, Saarbrücken, Germany) for his help with the thermal evaporation, and Dipti Rani for her assistance during prototyping. We acknowledge funding via NanoMatFutur grant of the German Ministry of Education and Research (FKZ13N13547). This work has received funding from the European Union’s Horizon 2020 research and innovation program under the Marie Skłodowska-Curie grant agreement N$^\circ$“765267”(QuSCo).

\textbf{Authors contribution:} E.N. acquired funding. O.R.O, R.N and E.N. planned the design, fabrication and the process development. O.R.O and R.N. worked on numerical simulation and optimization. O.R.O fabricated the antennas. O.R.O and R.N performed the characterization experiments. O.R.O and N.O. performed the application based experiments. R.N. and E.N supervised the project. O.R.O, N.O. and R.N. wrote the original draft. All the authors contributed during reviewing and editing.

The authors declare no conflict of interest.

\bibliography{References}

\begin{thebibliography}{65}%
\makeatletter
\providecommand \@ifxundefined [1]{%
 \@ifx{#1\undefined}
}%
\providecommand \@ifnum [1]{%
 \ifnum #1\expandafter \@firstoftwo
 \else \expandafter \@secondoftwo
 \fi
}%
\providecommand \@ifx [1]{%
 \ifx #1\expandafter \@firstoftwo
 \else \expandafter \@secondoftwo
 \fi
}%
\providecommand \natexlab [1]{#1}%
\providecommand \enquote  [1]{``#1''}%
\providecommand \bibnamefont  [1]{#1}%
\providecommand \bibfnamefont [1]{#1}%
\providecommand \citenamefont [1]{#1}%
\providecommand \href@noop [0]{\@secondoftwo}%
\providecommand \href [0]{\begingroup \@sanitize@url \@href}%
\providecommand \@href[1]{\@@startlink{#1}\@@href}%
\providecommand \@@href[1]{\endgroup#1\@@endlink}%
\providecommand \@sanitize@url [0]{\catcode `\\12\catcode `\$12\catcode
  `\&12\catcode `\#12\catcode `\^12\catcode `\_12\catcode `\%12\relax}%
\providecommand \@@startlink[1]{}%
\providecommand \@@endlink[0]{}%
\providecommand \url  [0]{\begingroup\@sanitize@url \@url }%
\providecommand \@url [1]{\endgroup\@href {#1}{\urlprefix }}%
\providecommand \urlprefix  [0]{URL }%
\providecommand \Eprint [0]{\href }%
\providecommand \doibase [0]{http://dx.doi.org/}%
\providecommand \selectlanguage [0]{\@gobble}%
\providecommand \bibinfo  [0]{\@secondoftwo}%
\providecommand \bibfield  [0]{\@secondoftwo}%
\providecommand \translation [1]{[#1]}%
\providecommand \BibitemOpen [0]{}%
\providecommand \bibitemStop [0]{}%
\providecommand \bibitemNoStop [0]{.\EOS\space}%
\providecommand \EOS [0]{\spacefactor3000\relax}%
\providecommand \BibitemShut  [1]{\csname bibitem#1\endcsname}%
\let\auto@bib@innerbib\@empty
\bibitem [{\citenamefont {Jaskula}\ \emph {et~al.}(2017)\citenamefont
  {Jaskula}, \citenamefont {Bauch}, \citenamefont {Arroyo-Camejo},
  \citenamefont {Lukin}, \citenamefont {Hell}, \citenamefont {Trifonov},\ and\
  \citenamefont {Walsworth}}]{jaskula2017superresolution}%
  \BibitemOpen
  \bibfield  {author} {\bibinfo {author} {\bibfnamefont {J.-C.}\ \bibnamefont
  {Jaskula}}, \bibinfo {author} {\bibfnamefont {E.}~\bibnamefont {Bauch}},
  \bibinfo {author} {\bibfnamefont {S.}~\bibnamefont {Arroyo-Camejo}}, \bibinfo
  {author} {\bibfnamefont {M.~D.}\ \bibnamefont {Lukin}}, \bibinfo {author}
  {\bibfnamefont {S.~W.}\ \bibnamefont {Hell}}, \bibinfo {author}
  {\bibfnamefont {A.~S.}\ \bibnamefont {Trifonov}}, \ and\ \bibinfo {author}
  {\bibfnamefont {R.~L.}\ \bibnamefont {Walsworth}},\ }\href@noop {} {\bibfield
   {journal} {\bibinfo  {journal} {Optics express}\ }\textbf {\bibinfo {volume}
  {25}},\ \bibinfo {pages} {11048} (\bibinfo {year} {2017})}\BibitemShut
  {NoStop}%
\bibitem [{\citenamefont {Barson}\ \emph {et~al.}(2017)\citenamefont {Barson},
  \citenamefont {Peddibhotla}, \citenamefont {Ovartchaiyapong}, \citenamefont
  {Ganesan}, \citenamefont {Taylor}, \citenamefont {Gebert}, \citenamefont
  {Mielens}, \citenamefont {Koslowski}, \citenamefont {Simpson}, \citenamefont
  {McGuinness} \emph {et~al.}}]{barson2017nanomechanical}%
  \BibitemOpen
  \bibfield  {author} {\bibinfo {author} {\bibfnamefont {M.~S.}\ \bibnamefont
  {Barson}}, \bibinfo {author} {\bibfnamefont {P.}~\bibnamefont {Peddibhotla}},
  \bibinfo {author} {\bibfnamefont {P.}~\bibnamefont {Ovartchaiyapong}},
  \bibinfo {author} {\bibfnamefont {K.}~\bibnamefont {Ganesan}}, \bibinfo
  {author} {\bibfnamefont {R.~L.}\ \bibnamefont {Taylor}}, \bibinfo {author}
  {\bibfnamefont {M.}~\bibnamefont {Gebert}}, \bibinfo {author} {\bibfnamefont
  {Z.}~\bibnamefont {Mielens}}, \bibinfo {author} {\bibfnamefont
  {B.}~\bibnamefont {Koslowski}}, \bibinfo {author} {\bibfnamefont {D.~A.}\
  \bibnamefont {Simpson}}, \bibinfo {author} {\bibfnamefont {L.~P.}\
  \bibnamefont {McGuinness}},  \emph {et~al.},\ }\href@noop {} {\bibfield
  {journal} {\bibinfo  {journal} {Nano letters}\ }\textbf {\bibinfo {volume}
  {17}},\ \bibinfo {pages} {1496} (\bibinfo {year} {2017})}\BibitemShut
  {NoStop}%
\bibitem [{\citenamefont {Pham}\ \emph {et~al.}(2016)\citenamefont {Pham},
  \citenamefont {DeVience}, \citenamefont {Casola}, \citenamefont {Lovchinsky},
  \citenamefont {Sushkov}, \citenamefont {Bersin}, \citenamefont {Lee},
  \citenamefont {Urbach}, \citenamefont {Cappellaro}, \citenamefont {Park}
  \emph {et~al.}}]{pham2016nmr}%
  \BibitemOpen
  \bibfield  {author} {\bibinfo {author} {\bibfnamefont {L.~M.}\ \bibnamefont
  {Pham}}, \bibinfo {author} {\bibfnamefont {S.~J.}\ \bibnamefont {DeVience}},
  \bibinfo {author} {\bibfnamefont {F.}~\bibnamefont {Casola}}, \bibinfo
  {author} {\bibfnamefont {I.}~\bibnamefont {Lovchinsky}}, \bibinfo {author}
  {\bibfnamefont {A.~O.}\ \bibnamefont {Sushkov}}, \bibinfo {author}
  {\bibfnamefont {E.}~\bibnamefont {Bersin}}, \bibinfo {author} {\bibfnamefont
  {J.}~\bibnamefont {Lee}}, \bibinfo {author} {\bibfnamefont {E.}~\bibnamefont
  {Urbach}}, \bibinfo {author} {\bibfnamefont {P.}~\bibnamefont {Cappellaro}},
  \bibinfo {author} {\bibfnamefont {H.}~\bibnamefont {Park}},  \emph {et~al.},\
  }\href@noop {} {\bibfield  {journal} {\bibinfo  {journal} {Physical Review
  B}\ }\textbf {\bibinfo {volume} {93}},\ \bibinfo {pages} {045425} (\bibinfo
  {year} {2016})}\BibitemShut {NoStop}%
\bibitem [{\citenamefont {Arai}\ \emph {et~al.}(2015)\citenamefont {Arai},
  \citenamefont {Belthangady}, \citenamefont {Zhang}, \citenamefont {Bar-Gill},
  \citenamefont {DeVience}, \citenamefont {Cappellaro}, \citenamefont
  {Yacoby},\ and\ \citenamefont {Walsworth}}]{arai2015fourier}%
  \BibitemOpen
  \bibfield  {author} {\bibinfo {author} {\bibfnamefont {K.}~\bibnamefont
  {Arai}}, \bibinfo {author} {\bibfnamefont {C.}~\bibnamefont {Belthangady}},
  \bibinfo {author} {\bibfnamefont {H.}~\bibnamefont {Zhang}}, \bibinfo
  {author} {\bibfnamefont {N.}~\bibnamefont {Bar-Gill}}, \bibinfo {author}
  {\bibfnamefont {S.}~\bibnamefont {DeVience}}, \bibinfo {author}
  {\bibfnamefont {P.}~\bibnamefont {Cappellaro}}, \bibinfo {author}
  {\bibfnamefont {A.}~\bibnamefont {Yacoby}}, \ and\ \bibinfo {author}
  {\bibfnamefont {R.~L.}\ \bibnamefont {Walsworth}},\ }\href@noop {} {\bibfield
   {journal} {\bibinfo  {journal} {Nature nanotechnology}\ }\textbf {\bibinfo
  {volume} {10}},\ \bibinfo {pages} {859} (\bibinfo {year} {2015})}\BibitemShut
  {NoStop}%
\bibitem [{\citenamefont {Grinolds}\ \emph {et~al.}(2014)\citenamefont
  {Grinolds}, \citenamefont {Warner}, \citenamefont {De~Greve}, \citenamefont
  {Dovzhenko}, \citenamefont {Thiel}, \citenamefont {Walsworth}, \citenamefont
  {Hong}, \citenamefont {Maletinsky},\ and\ \citenamefont
  {Yacoby}}]{grinolds2014subnanometre}%
  \BibitemOpen
  \bibfield  {author} {\bibinfo {author} {\bibfnamefont {M.}~\bibnamefont
  {Grinolds}}, \bibinfo {author} {\bibfnamefont {M.}~\bibnamefont {Warner}},
  \bibinfo {author} {\bibfnamefont {K.}~\bibnamefont {De~Greve}}, \bibinfo
  {author} {\bibfnamefont {Y.}~\bibnamefont {Dovzhenko}}, \bibinfo {author}
  {\bibfnamefont {L.}~\bibnamefont {Thiel}}, \bibinfo {author} {\bibfnamefont
  {R.~L.}\ \bibnamefont {Walsworth}}, \bibinfo {author} {\bibfnamefont
  {S.}~\bibnamefont {Hong}}, \bibinfo {author} {\bibfnamefont {P.}~\bibnamefont
  {Maletinsky}}, \ and\ \bibinfo {author} {\bibfnamefont {A.}~\bibnamefont
  {Yacoby}},\ }\href@noop {} {\bibfield  {journal} {\bibinfo  {journal} {Nature
  nanotechnology}\ }\textbf {\bibinfo {volume} {9}},\ \bibinfo {pages} {279}
  (\bibinfo {year} {2014})}\BibitemShut {NoStop}%
\bibitem [{\citenamefont {Doherty}\ \emph {et~al.}(2014)\citenamefont
  {Doherty}, \citenamefont {Struzhkin}, \citenamefont {Simpson}, \citenamefont
  {McGuinness}, \citenamefont {Meng}, \citenamefont {Stacey}, \citenamefont
  {Karle}, \citenamefont {Hemley}, \citenamefont {Manson}, \citenamefont
  {Hollenberg} \emph {et~al.}}]{doherty2014electronic}%
  \BibitemOpen
  \bibfield  {author} {\bibinfo {author} {\bibfnamefont {M.~W.}\ \bibnamefont
  {Doherty}}, \bibinfo {author} {\bibfnamefont {V.~V.}\ \bibnamefont
  {Struzhkin}}, \bibinfo {author} {\bibfnamefont {D.~A.}\ \bibnamefont
  {Simpson}}, \bibinfo {author} {\bibfnamefont {L.~P.}\ \bibnamefont
  {McGuinness}}, \bibinfo {author} {\bibfnamefont {Y.}~\bibnamefont {Meng}},
  \bibinfo {author} {\bibfnamefont {A.}~\bibnamefont {Stacey}}, \bibinfo
  {author} {\bibfnamefont {T.~J.}\ \bibnamefont {Karle}}, \bibinfo {author}
  {\bibfnamefont {R.~J.}\ \bibnamefont {Hemley}}, \bibinfo {author}
  {\bibfnamefont {N.~B.}\ \bibnamefont {Manson}}, \bibinfo {author}
  {\bibfnamefont {L.~C.}\ \bibnamefont {Hollenberg}},  \emph {et~al.},\
  }\href@noop {} {\bibfield  {journal} {\bibinfo  {journal} {Physical review
  letters}\ }\textbf {\bibinfo {volume} {112}},\ \bibinfo {pages} {047601}
  (\bibinfo {year} {2014})}\BibitemShut {NoStop}%
\bibitem [{\citenamefont {Plakhotnik}\ \emph {et~al.}(2014)\citenamefont
  {Plakhotnik}, \citenamefont {Doherty}, \citenamefont {Cole}, \citenamefont
  {Chapman},\ and\ \citenamefont {Manson}}]{plakhotnik2014all}%
  \BibitemOpen
  \bibfield  {author} {\bibinfo {author} {\bibfnamefont {T.}~\bibnamefont
  {Plakhotnik}}, \bibinfo {author} {\bibfnamefont {M.~W.}\ \bibnamefont
  {Doherty}}, \bibinfo {author} {\bibfnamefont {J.~H.}\ \bibnamefont {Cole}},
  \bibinfo {author} {\bibfnamefont {R.}~\bibnamefont {Chapman}}, \ and\
  \bibinfo {author} {\bibfnamefont {N.~B.}\ \bibnamefont {Manson}},\
  }\href@noop {} {\bibfield  {journal} {\bibinfo  {journal} {Nano letters}\
  }\textbf {\bibinfo {volume} {14}},\ \bibinfo {pages} {4989} (\bibinfo {year}
  {2014})}\BibitemShut {NoStop}%
\bibitem [{\citenamefont {Kucsko}\ \emph {et~al.}(2013)\citenamefont {Kucsko},
  \citenamefont {Maurer}, \citenamefont {Yao}, \citenamefont {Kubo},
  \citenamefont {Noh}, \citenamefont {Lo}, \citenamefont {Park},\ and\
  \citenamefont {Lukin}}]{kucsko2013nanometre}%
  \BibitemOpen
  \bibfield  {author} {\bibinfo {author} {\bibfnamefont {G.}~\bibnamefont
  {Kucsko}}, \bibinfo {author} {\bibfnamefont {P.~C.}\ \bibnamefont {Maurer}},
  \bibinfo {author} {\bibfnamefont {N.~Y.}\ \bibnamefont {Yao}}, \bibinfo
  {author} {\bibfnamefont {M.}~\bibnamefont {Kubo}}, \bibinfo {author}
  {\bibfnamefont {H.~J.}\ \bibnamefont {Noh}}, \bibinfo {author} {\bibfnamefont
  {P.~K.}\ \bibnamefont {Lo}}, \bibinfo {author} {\bibfnamefont
  {H.}~\bibnamefont {Park}}, \ and\ \bibinfo {author} {\bibfnamefont {M.~D.}\
  \bibnamefont {Lukin}},\ }\href@noop {} {\bibfield  {journal} {\bibinfo
  {journal} {Nature}\ }\textbf {\bibinfo {volume} {500}},\ \bibinfo {pages}
  {54} (\bibinfo {year} {2013})}\BibitemShut {NoStop}%
\bibitem [{\citenamefont {Doherty}\ \emph {et~al.}(2013)\citenamefont
  {Doherty}, \citenamefont {Manson}, \citenamefont {Delaney}, \citenamefont
  {Jelezko}, \citenamefont {Wrachtrup},\ and\ \citenamefont
  {Hollenberg}}]{doherty2013nitrogen}%
  \BibitemOpen
  \bibfield  {author} {\bibinfo {author} {\bibfnamefont {M.~W.}\ \bibnamefont
  {Doherty}}, \bibinfo {author} {\bibfnamefont {N.~B.}\ \bibnamefont {Manson}},
  \bibinfo {author} {\bibfnamefont {P.}~\bibnamefont {Delaney}}, \bibinfo
  {author} {\bibfnamefont {F.}~\bibnamefont {Jelezko}}, \bibinfo {author}
  {\bibfnamefont {J.}~\bibnamefont {Wrachtrup}}, \ and\ \bibinfo {author}
  {\bibfnamefont {L.~C.}\ \bibnamefont {Hollenberg}},\ }\href@noop {}
  {\bibfield  {journal} {\bibinfo  {journal} {Physics Reports}\ }\textbf
  {\bibinfo {volume} {528}},\ \bibinfo {pages} {1} (\bibinfo {year}
  {2013})}\BibitemShut {NoStop}%
\bibitem [{\citenamefont {Goldman}\ \emph
  {et~al.}(2015{\natexlab{a}})\citenamefont {Goldman}, \citenamefont {Doherty},
  \citenamefont {Sipahigil}, \citenamefont {Yao}, \citenamefont {Bennett},
  \citenamefont {Manson}, \citenamefont {Kubanek},\ and\ \citenamefont
  {Lukin}}]{goldman2015state}%
  \BibitemOpen
  \bibfield  {author} {\bibinfo {author} {\bibfnamefont {M.~L.}\ \bibnamefont
  {Goldman}}, \bibinfo {author} {\bibfnamefont {M.}~\bibnamefont {Doherty}},
  \bibinfo {author} {\bibfnamefont {A.}~\bibnamefont {Sipahigil}}, \bibinfo
  {author} {\bibfnamefont {N.~Y.}\ \bibnamefont {Yao}}, \bibinfo {author}
  {\bibfnamefont {S.}~\bibnamefont {Bennett}}, \bibinfo {author} {\bibfnamefont
  {N.}~\bibnamefont {Manson}}, \bibinfo {author} {\bibfnamefont
  {A.}~\bibnamefont {Kubanek}}, \ and\ \bibinfo {author} {\bibfnamefont
  {M.~D.}\ \bibnamefont {Lukin}},\ }\href@noop {} {\bibfield  {journal}
  {\bibinfo  {journal} {Physical Review B}\ }\textbf {\bibinfo {volume} {91}},\
  \bibinfo {pages} {165201} (\bibinfo {year} {2015}{\natexlab{a}})}\BibitemShut
  {NoStop}%
\bibitem [{\citenamefont {Goldman}\ \emph
  {et~al.}(2015{\natexlab{b}})\citenamefont {Goldman}, \citenamefont
  {Sipahigil}, \citenamefont {Doherty}, \citenamefont {Yao}, \citenamefont
  {Bennett}, \citenamefont {Markham}, \citenamefont {Twitchen}, \citenamefont
  {Manson}, \citenamefont {Kubanek},\ and\ \citenamefont
  {Lukin}}]{goldman2015phonon}%
  \BibitemOpen
  \bibfield  {author} {\bibinfo {author} {\bibfnamefont {M.~L.}\ \bibnamefont
  {Goldman}}, \bibinfo {author} {\bibfnamefont {A.}~\bibnamefont {Sipahigil}},
  \bibinfo {author} {\bibfnamefont {M.}~\bibnamefont {Doherty}}, \bibinfo
  {author} {\bibfnamefont {N.~Y.}\ \bibnamefont {Yao}}, \bibinfo {author}
  {\bibfnamefont {S.}~\bibnamefont {Bennett}}, \bibinfo {author} {\bibfnamefont
  {M.}~\bibnamefont {Markham}}, \bibinfo {author} {\bibfnamefont
  {D.}~\bibnamefont {Twitchen}}, \bibinfo {author} {\bibfnamefont
  {N.}~\bibnamefont {Manson}}, \bibinfo {author} {\bibfnamefont
  {A.}~\bibnamefont {Kubanek}}, \ and\ \bibinfo {author} {\bibfnamefont
  {M.~D.}\ \bibnamefont {Lukin}},\ }\href@noop {} {\bibfield  {journal}
  {\bibinfo  {journal} {Physical review letters}\ }\textbf {\bibinfo {volume}
  {114}},\ \bibinfo {pages} {145502} (\bibinfo {year}
  {2015}{\natexlab{b}})}\BibitemShut {NoStop}%
\bibitem [{\citenamefont {Barry}\ \emph {et~al.}(2020)\citenamefont {Barry},
  \citenamefont {Schloss}, \citenamefont {Bauch}, \citenamefont {Turner},
  \citenamefont {Hart}, \citenamefont {Pham},\ and\ \citenamefont
  {Walsworth}}]{Barry2020}%
  \BibitemOpen
  \bibfield  {author} {\bibinfo {author} {\bibfnamefont {J.~F.}\ \bibnamefont
  {Barry}}, \bibinfo {author} {\bibfnamefont {J.~M.}\ \bibnamefont {Schloss}},
  \bibinfo {author} {\bibfnamefont {E.}~\bibnamefont {Bauch}}, \bibinfo
  {author} {\bibfnamefont {M.~J.}\ \bibnamefont {Turner}}, \bibinfo {author}
  {\bibfnamefont {C.~A.}\ \bibnamefont {Hart}}, \bibinfo {author}
  {\bibfnamefont {L.~M.}\ \bibnamefont {Pham}}, \ and\ \bibinfo {author}
  {\bibfnamefont {R.~L.}\ \bibnamefont {Walsworth}},\ }\href {\doibase
  10.1103/RevModPhys.92.015004} {\bibfield  {journal} {\bibinfo  {journal}
  {Rev. Mod. Phys.}\ }\textbf {\bibinfo {volume} {92}},\ \bibinfo {pages}
  {015004} (\bibinfo {year} {2020})}\BibitemShut {NoStop}%
\bibitem [{\citenamefont {Bernardi}\ \emph {et~al.}(2017)\citenamefont
  {Bernardi}, \citenamefont {Nelz}, \citenamefont {Sonusen},\ and\
  \citenamefont {Neu}}]{bernardi2017nanoscale}%
  \BibitemOpen
  \bibfield  {author} {\bibinfo {author} {\bibfnamefont {E.}~\bibnamefont
  {Bernardi}}, \bibinfo {author} {\bibfnamefont {R.}~\bibnamefont {Nelz}},
  \bibinfo {author} {\bibfnamefont {S.}~\bibnamefont {Sonusen}}, \ and\
  \bibinfo {author} {\bibfnamefont {E.}~\bibnamefont {Neu}},\ }\href@noop {}
  {\bibfield  {journal} {\bibinfo  {journal} {Crystals}\ }\textbf {\bibinfo
  {volume} {7}},\ \bibinfo {pages} {124} (\bibinfo {year} {2017})}\BibitemShut
  {NoStop}%
\bibitem [{\citenamefont {Hahn}(1950)}]{hahn1950spin}%
  \BibitemOpen
  \bibfield  {author} {\bibinfo {author} {\bibfnamefont {E.~L.}\ \bibnamefont
  {Hahn}},\ }\href@noop {} {\bibfield  {journal} {\bibinfo  {journal} {Physical
  review}\ }\textbf {\bibinfo {volume} {80}},\ \bibinfo {pages} {580} (\bibinfo
  {year} {1950})}\BibitemShut {NoStop}%
\bibitem [{\citenamefont {Carr}\ and\ \citenamefont
  {Purcell}(1954)}]{Carr1954}%
  \BibitemOpen
  \bibfield  {author} {\bibinfo {author} {\bibfnamefont {H.~Y.}\ \bibnamefont
  {Carr}}\ and\ \bibinfo {author} {\bibfnamefont {E.~M.}\ \bibnamefont
  {Purcell}},\ }\href {\doibase 10.1103/PhysRev.94.630} {\bibfield  {journal}
  {\bibinfo  {journal} {Phys. Rev.}\ }\textbf {\bibinfo {volume} {94}},\
  \bibinfo {pages} {630} (\bibinfo {year} {1954})}\BibitemShut {NoStop}%
\bibitem [{\citenamefont {Gullion}\ \emph {et~al.}(1990)\citenamefont
  {Gullion}, \citenamefont {Baker},\ and\ \citenamefont
  {Conradi}}]{gullion1990new}%
  \BibitemOpen
  \bibfield  {author} {\bibinfo {author} {\bibfnamefont {T.}~\bibnamefont
  {Gullion}}, \bibinfo {author} {\bibfnamefont {D.~B.}\ \bibnamefont {Baker}},
  \ and\ \bibinfo {author} {\bibfnamefont {M.~S.}\ \bibnamefont {Conradi}},\
  }\href@noop {} {\bibfield  {journal} {\bibinfo  {journal} {Journal of
  Magnetic Resonance (1969)}\ }\textbf {\bibinfo {volume} {89}},\ \bibinfo
  {pages} {479} (\bibinfo {year} {1990})}\BibitemShut {NoStop}%
\bibitem [{\citenamefont {Viola}\ and\ \citenamefont
  {Lloyd}(1998)}]{viola1998dynamical}%
  \BibitemOpen
  \bibfield  {author} {\bibinfo {author} {\bibfnamefont {L.}~\bibnamefont
  {Viola}}\ and\ \bibinfo {author} {\bibfnamefont {S.}~\bibnamefont {Lloyd}},\
  }\href@noop {} {\bibfield  {journal} {\bibinfo  {journal} {Physical Review
  A}\ }\textbf {\bibinfo {volume} {58}},\ \bibinfo {pages} {2733} (\bibinfo
  {year} {1998})}\BibitemShut {NoStop}%
\bibitem [{\citenamefont {Cywi{\'n}ski}\ \emph {et~al.}(2008)\citenamefont
  {Cywi{\'n}ski}, \citenamefont {Lutchyn}, \citenamefont {Nave},\ and\
  \citenamefont {Sarma}}]{cywinski2008enhance}%
  \BibitemOpen
  \bibfield  {author} {\bibinfo {author} {\bibfnamefont {{\L}.}~\bibnamefont
  {Cywi{\'n}ski}}, \bibinfo {author} {\bibfnamefont {R.~M.}\ \bibnamefont
  {Lutchyn}}, \bibinfo {author} {\bibfnamefont {C.~P.}\ \bibnamefont {Nave}}, \
  and\ \bibinfo {author} {\bibfnamefont {S.~D.}\ \bibnamefont {Sarma}},\
  }\href@noop {} {\bibfield  {journal} {\bibinfo  {journal} {Physical Review
  B}\ }\textbf {\bibinfo {volume} {77}},\ \bibinfo {pages} {174509} (\bibinfo
  {year} {2008})}\BibitemShut {NoStop}%
\bibitem [{\citenamefont {Slichter}(2013)}]{slichter2013principles}%
  \BibitemOpen
  \bibfield  {author} {\bibinfo {author} {\bibfnamefont {C.~P.}\ \bibnamefont
  {Slichter}},\ }\href@noop {} {\emph {\bibinfo {title} {Principles of magnetic
  resonance}}},\ Vol.~\bibinfo {volume} {1}\ (\bibinfo  {publisher} {Springer
  Science \& Business Media},\ \bibinfo {year} {2013})\BibitemShut {NoStop}%
\bibitem [{\citenamefont {Degen}\ \emph {et~al.}(2017)\citenamefont {Degen},
  \citenamefont {Reinhard},\ and\ \citenamefont
  {Cappellaro}}]{degen2017quantum}%
  \BibitemOpen
  \bibfield  {author} {\bibinfo {author} {\bibfnamefont {C.~L.}\ \bibnamefont
  {Degen}}, \bibinfo {author} {\bibfnamefont {F.}~\bibnamefont {Reinhard}}, \
  and\ \bibinfo {author} {\bibfnamefont {P.}~\bibnamefont {Cappellaro}},\
  }\href {\doibase 10.1103/RevModPhys.89.035002} {\bibfield  {journal}
  {\bibinfo  {journal} {Reviews of modern physics}\ }\textbf {\bibinfo {volume}
  {89}},\ \bibinfo {pages} {035002} (\bibinfo {year} {2017})}\BibitemShut
  {NoStop}%
\bibitem [{\citenamefont {Turner}\ \emph {et~al.}(2020)\citenamefont {Turner},
  \citenamefont {Langellier}, \citenamefont {Bainbridge}, \citenamefont
  {Walters}, \citenamefont {Meesala}, \citenamefont {Babinec}, \citenamefont
  {Kehayias}, \citenamefont {Yacoby}, \citenamefont {Hu}, \citenamefont
  {Lon{\v{c}}ar} \emph {et~al.}}]{turner2020magnetic}%
  \BibitemOpen
  \bibfield  {author} {\bibinfo {author} {\bibfnamefont {M.~J.}\ \bibnamefont
  {Turner}}, \bibinfo {author} {\bibfnamefont {N.}~\bibnamefont {Langellier}},
  \bibinfo {author} {\bibfnamefont {R.}~\bibnamefont {Bainbridge}}, \bibinfo
  {author} {\bibfnamefont {D.}~\bibnamefont {Walters}}, \bibinfo {author}
  {\bibfnamefont {S.}~\bibnamefont {Meesala}}, \bibinfo {author} {\bibfnamefont
  {T.~M.}\ \bibnamefont {Babinec}}, \bibinfo {author} {\bibfnamefont
  {P.}~\bibnamefont {Kehayias}}, \bibinfo {author} {\bibfnamefont
  {A.}~\bibnamefont {Yacoby}}, \bibinfo {author} {\bibfnamefont
  {E.}~\bibnamefont {Hu}}, \bibinfo {author} {\bibfnamefont {M.}~\bibnamefont
  {Lon{\v{c}}ar}},  \emph {et~al.},\ }\href@noop {} {\bibfield  {journal}
  {\bibinfo  {journal} {Physical Review Applied}\ }\textbf {\bibinfo {volume}
  {14}},\ \bibinfo {pages} {014097} (\bibinfo {year} {2020})}\BibitemShut
  {NoStop}%
\bibitem [{\citenamefont {Mizuno}\ \emph {et~al.}(2020)\citenamefont {Mizuno},
  \citenamefont {Ishiwata}, \citenamefont {Masuyama}, \citenamefont {Iwasaki},\
  and\ \citenamefont {Hatano}}]{mizuno2020simultaneous}%
  \BibitemOpen
  \bibfield  {author} {\bibinfo {author} {\bibfnamefont {K.}~\bibnamefont
  {Mizuno}}, \bibinfo {author} {\bibfnamefont {H.}~\bibnamefont {Ishiwata}},
  \bibinfo {author} {\bibfnamefont {Y.}~\bibnamefont {Masuyama}}, \bibinfo
  {author} {\bibfnamefont {T.}~\bibnamefont {Iwasaki}}, \ and\ \bibinfo
  {author} {\bibfnamefont {M.}~\bibnamefont {Hatano}},\ }\href@noop {}
  {\bibfield  {journal} {\bibinfo  {journal} {Scientific Reports}\ }\textbf
  {\bibinfo {volume} {10}},\ \bibinfo {pages} {1} (\bibinfo {year}
  {2020})}\BibitemShut {NoStop}%
\bibitem [{\citenamefont {Horsley}\ \emph {et~al.}(2018)\citenamefont
  {Horsley}, \citenamefont {Appel}, \citenamefont {Wolters}, \citenamefont
  {Achard}, \citenamefont {Tallaire}, \citenamefont {Maletinsky},\ and\
  \citenamefont {Treutlein}}]{horsley2018microwave}%
  \BibitemOpen
  \bibfield  {author} {\bibinfo {author} {\bibfnamefont {A.}~\bibnamefont
  {Horsley}}, \bibinfo {author} {\bibfnamefont {P.}~\bibnamefont {Appel}},
  \bibinfo {author} {\bibfnamefont {J.}~\bibnamefont {Wolters}}, \bibinfo
  {author} {\bibfnamefont {J.}~\bibnamefont {Achard}}, \bibinfo {author}
  {\bibfnamefont {A.}~\bibnamefont {Tallaire}}, \bibinfo {author}
  {\bibfnamefont {P.}~\bibnamefont {Maletinsky}}, \ and\ \bibinfo {author}
  {\bibfnamefont {P.}~\bibnamefont {Treutlein}},\ }\href@noop {} {\bibfield
  {journal} {\bibinfo  {journal} {Physical Review Applied}\ }\textbf {\bibinfo
  {volume} {10}},\ \bibinfo {pages} {044039} (\bibinfo {year}
  {2018})}\BibitemShut {NoStop}%
\bibitem [{\citenamefont {Glenn}\ \emph {et~al.}(2017)\citenamefont {Glenn},
  \citenamefont {Fu}, \citenamefont {Kehayias}, \citenamefont {Le~Sage},
  \citenamefont {Lima}, \citenamefont {Weiss},\ and\ \citenamefont
  {Walsworth}}]{glenn2017micrometer}%
  \BibitemOpen
  \bibfield  {author} {\bibinfo {author} {\bibfnamefont {D.~R.}\ \bibnamefont
  {Glenn}}, \bibinfo {author} {\bibfnamefont {R.~R.}\ \bibnamefont {Fu}},
  \bibinfo {author} {\bibfnamefont {P.}~\bibnamefont {Kehayias}}, \bibinfo
  {author} {\bibfnamefont {D.}~\bibnamefont {Le~Sage}}, \bibinfo {author}
  {\bibfnamefont {E.~A.}\ \bibnamefont {Lima}}, \bibinfo {author}
  {\bibfnamefont {B.~P.}\ \bibnamefont {Weiss}}, \ and\ \bibinfo {author}
  {\bibfnamefont {R.~L.}\ \bibnamefont {Walsworth}},\ }\href@noop {} {\bibfield
   {journal} {\bibinfo  {journal} {Geochemistry, Geophysics, Geosystems}\
  }\textbf {\bibinfo {volume} {18}},\ \bibinfo {pages} {3254} (\bibinfo {year}
  {2017})}\BibitemShut {NoStop}%
\bibitem [{\citenamefont {Simpson}\ \emph {et~al.}(2016)\citenamefont
  {Simpson}, \citenamefont {Tetienne}, \citenamefont {McCoey}, \citenamefont
  {Ganesan}, \citenamefont {Hall}, \citenamefont {Petrou}, \citenamefont
  {Scholten},\ and\ \citenamefont {Hollenberg}}]{simpson2016magneto}%
  \BibitemOpen
  \bibfield  {author} {\bibinfo {author} {\bibfnamefont {D.~A.}\ \bibnamefont
  {Simpson}}, \bibinfo {author} {\bibfnamefont {J.-P.}\ \bibnamefont
  {Tetienne}}, \bibinfo {author} {\bibfnamefont {J.~M.}\ \bibnamefont
  {McCoey}}, \bibinfo {author} {\bibfnamefont {K.}~\bibnamefont {Ganesan}},
  \bibinfo {author} {\bibfnamefont {L.~T.}\ \bibnamefont {Hall}}, \bibinfo
  {author} {\bibfnamefont {S.}~\bibnamefont {Petrou}}, \bibinfo {author}
  {\bibfnamefont {R.~E.}\ \bibnamefont {Scholten}}, \ and\ \bibinfo {author}
  {\bibfnamefont {L.~C.}\ \bibnamefont {Hollenberg}},\ }\href@noop {}
  {\bibfield  {journal} {\bibinfo  {journal} {Scientific reports}\ }\textbf
  {\bibinfo {volume} {6}},\ \bibinfo {pages} {1} (\bibinfo {year}
  {2016})}\BibitemShut {NoStop}%
\bibitem [{\citenamefont {Clevenson}\ \emph {et~al.}(2015)\citenamefont
  {Clevenson}, \citenamefont {Trusheim}, \citenamefont {Teale}, \citenamefont
  {Schr{\"o}der}, \citenamefont {Braje},\ and\ \citenamefont
  {Englund}}]{clevenson2015broadband}%
  \BibitemOpen
  \bibfield  {author} {\bibinfo {author} {\bibfnamefont {H.}~\bibnamefont
  {Clevenson}}, \bibinfo {author} {\bibfnamefont {M.~E.}\ \bibnamefont
  {Trusheim}}, \bibinfo {author} {\bibfnamefont {C.}~\bibnamefont {Teale}},
  \bibinfo {author} {\bibfnamefont {T.}~\bibnamefont {Schr{\"o}der}}, \bibinfo
  {author} {\bibfnamefont {D.}~\bibnamefont {Braje}}, \ and\ \bibinfo {author}
  {\bibfnamefont {D.}~\bibnamefont {Englund}},\ }\href@noop {} {\bibfield
  {journal} {\bibinfo  {journal} {Nature Physics}\ }\textbf {\bibinfo {volume}
  {11}},\ \bibinfo {pages} {393} (\bibinfo {year} {2015})}\BibitemShut
  {NoStop}%
\bibitem [{\citenamefont {Nowodzinski}\ \emph {et~al.}(2015)\citenamefont
  {Nowodzinski}, \citenamefont {Chipaux}, \citenamefont {Toraille},
  \citenamefont {Jacques}, \citenamefont {Roch},\ and\ \citenamefont
  {Debuisschert}}]{nowodzinski2015nitrogen}%
  \BibitemOpen
  \bibfield  {author} {\bibinfo {author} {\bibfnamefont {A.}~\bibnamefont
  {Nowodzinski}}, \bibinfo {author} {\bibfnamefont {M.}~\bibnamefont
  {Chipaux}}, \bibinfo {author} {\bibfnamefont {L.}~\bibnamefont {Toraille}},
  \bibinfo {author} {\bibfnamefont {V.}~\bibnamefont {Jacques}}, \bibinfo
  {author} {\bibfnamefont {J.-F.}\ \bibnamefont {Roch}}, \ and\ \bibinfo
  {author} {\bibfnamefont {T.}~\bibnamefont {Debuisschert}},\ }\href@noop {}
  {\bibfield  {journal} {\bibinfo  {journal} {Microelectronics Reliability}\
  }\textbf {\bibinfo {volume} {55}},\ \bibinfo {pages} {1549} (\bibinfo {year}
  {2015})}\BibitemShut {NoStop}%
\bibitem [{\citenamefont {Chipaux}\ \emph {et~al.}(2015)\citenamefont
  {Chipaux}, \citenamefont {Tallaire}, \citenamefont {Achard}, \citenamefont
  {Pezzagna}, \citenamefont {Meijer}, \citenamefont {Jacques}, \citenamefont
  {Roch},\ and\ \citenamefont {Debuisschert}}]{chipaux2015magnetic}%
  \BibitemOpen
  \bibfield  {author} {\bibinfo {author} {\bibfnamefont {M.}~\bibnamefont
  {Chipaux}}, \bibinfo {author} {\bibfnamefont {A.}~\bibnamefont {Tallaire}},
  \bibinfo {author} {\bibfnamefont {J.}~\bibnamefont {Achard}}, \bibinfo
  {author} {\bibfnamefont {S.}~\bibnamefont {Pezzagna}}, \bibinfo {author}
  {\bibfnamefont {J.}~\bibnamefont {Meijer}}, \bibinfo {author} {\bibfnamefont
  {V.}~\bibnamefont {Jacques}}, \bibinfo {author} {\bibfnamefont {J.-F.}\
  \bibnamefont {Roch}}, \ and\ \bibinfo {author} {\bibfnamefont
  {T.}~\bibnamefont {Debuisschert}},\ }\href@noop {} {\bibfield  {journal}
  {\bibinfo  {journal} {The European Physical Journal D}\ }\textbf {\bibinfo
  {volume} {69}},\ \bibinfo {pages} {1} (\bibinfo {year} {2015})}\BibitemShut
  {NoStop}%
\bibitem [{\citenamefont {Le~Sage}\ \emph {et~al.}(2013)\citenamefont
  {Le~Sage}, \citenamefont {Arai}, \citenamefont {Glenn}, \citenamefont
  {DeVience}, \citenamefont {Pham}, \citenamefont {Rahn-Lee}, \citenamefont
  {Lukin}, \citenamefont {Yacoby}, \citenamefont {Komeili},\ and\ \citenamefont
  {Walsworth}}]{le2013optical}%
  \BibitemOpen
  \bibfield  {author} {\bibinfo {author} {\bibfnamefont {D.}~\bibnamefont
  {Le~Sage}}, \bibinfo {author} {\bibfnamefont {K.}~\bibnamefont {Arai}},
  \bibinfo {author} {\bibfnamefont {D.~R.}\ \bibnamefont {Glenn}}, \bibinfo
  {author} {\bibfnamefont {S.~J.}\ \bibnamefont {DeVience}}, \bibinfo {author}
  {\bibfnamefont {L.~M.}\ \bibnamefont {Pham}}, \bibinfo {author}
  {\bibfnamefont {L.}~\bibnamefont {Rahn-Lee}}, \bibinfo {author}
  {\bibfnamefont {M.~D.}\ \bibnamefont {Lukin}}, \bibinfo {author}
  {\bibfnamefont {A.}~\bibnamefont {Yacoby}}, \bibinfo {author} {\bibfnamefont
  {A.}~\bibnamefont {Komeili}}, \ and\ \bibinfo {author} {\bibfnamefont
  {R.~L.}\ \bibnamefont {Walsworth}},\ }\href@noop {} {\bibfield  {journal}
  {\bibinfo  {journal} {Nature}\ }\textbf {\bibinfo {volume} {496}},\ \bibinfo
  {pages} {486} (\bibinfo {year} {2013})}\BibitemShut {NoStop}%
\bibitem [{\citenamefont {Steinert}\ \emph {et~al.}(2013)\citenamefont
  {Steinert}, \citenamefont {Ziem}, \citenamefont {Hall}, \citenamefont
  {Zappe}, \citenamefont {Schweikert}, \citenamefont {G{\"o}tz}, \citenamefont
  {Aird}, \citenamefont {Balasubramanian}, \citenamefont {Hollenberg},\ and\
  \citenamefont {Wrachtrup}}]{steinert2013magnetic}%
  \BibitemOpen
  \bibfield  {author} {\bibinfo {author} {\bibfnamefont {S.}~\bibnamefont
  {Steinert}}, \bibinfo {author} {\bibfnamefont {F.}~\bibnamefont {Ziem}},
  \bibinfo {author} {\bibfnamefont {L.}~\bibnamefont {Hall}}, \bibinfo {author}
  {\bibfnamefont {A.}~\bibnamefont {Zappe}}, \bibinfo {author} {\bibfnamefont
  {M.}~\bibnamefont {Schweikert}}, \bibinfo {author} {\bibfnamefont
  {N.}~\bibnamefont {G{\"o}tz}}, \bibinfo {author} {\bibfnamefont
  {A.}~\bibnamefont {Aird}}, \bibinfo {author} {\bibfnamefont {G.}~\bibnamefont
  {Balasubramanian}}, \bibinfo {author} {\bibfnamefont {L.}~\bibnamefont
  {Hollenberg}}, \ and\ \bibinfo {author} {\bibfnamefont {J.}~\bibnamefont
  {Wrachtrup}},\ }\href@noop {} {\bibfield  {journal} {\bibinfo  {journal}
  {Nature communications}\ }\textbf {\bibinfo {volume} {4}},\ \bibinfo {pages}
  {1} (\bibinfo {year} {2013})}\BibitemShut {NoStop}%
\bibitem [{\citenamefont {Pham}\ \emph {et~al.}(2011)\citenamefont {Pham},
  \citenamefont {Le~Sage}, \citenamefont {Stanwix}, \citenamefont {Yeung},
  \citenamefont {Glenn}, \citenamefont {Trifonov}, \citenamefont {Cappellaro},
  \citenamefont {Hemmer}, \citenamefont {Lukin}, \citenamefont {Park} \emph
  {et~al.}}]{pham2011magnetic}%
  \BibitemOpen
  \bibfield  {author} {\bibinfo {author} {\bibfnamefont {L.~M.}\ \bibnamefont
  {Pham}}, \bibinfo {author} {\bibfnamefont {D.}~\bibnamefont {Le~Sage}},
  \bibinfo {author} {\bibfnamefont {P.~L.}\ \bibnamefont {Stanwix}}, \bibinfo
  {author} {\bibfnamefont {T.~K.}\ \bibnamefont {Yeung}}, \bibinfo {author}
  {\bibfnamefont {D.}~\bibnamefont {Glenn}}, \bibinfo {author} {\bibfnamefont
  {A.}~\bibnamefont {Trifonov}}, \bibinfo {author} {\bibfnamefont
  {P.}~\bibnamefont {Cappellaro}}, \bibinfo {author} {\bibfnamefont {P.~R.}\
  \bibnamefont {Hemmer}}, \bibinfo {author} {\bibfnamefont {M.~D.}\
  \bibnamefont {Lukin}}, \bibinfo {author} {\bibfnamefont {H.}~\bibnamefont
  {Park}},  \emph {et~al.},\ }\href@noop {} {\bibfield  {journal} {\bibinfo
  {journal} {New Journal of Physics}\ }\textbf {\bibinfo {volume} {13}},\
  \bibinfo {pages} {045021} (\bibinfo {year} {2011})}\BibitemShut {NoStop}%
\bibitem [{\citenamefont {Steinert}\ \emph {et~al.}(2010)\citenamefont
  {Steinert}, \citenamefont {Dolde}, \citenamefont {Neumann}, \citenamefont
  {Aird}, \citenamefont {Naydenov}, \citenamefont {Balasubramanian},
  \citenamefont {Jelezko},\ and\ \citenamefont {Wrachtrup}}]{steinert2010high}%
  \BibitemOpen
  \bibfield  {author} {\bibinfo {author} {\bibfnamefont {S.}~\bibnamefont
  {Steinert}}, \bibinfo {author} {\bibfnamefont {F.}~\bibnamefont {Dolde}},
  \bibinfo {author} {\bibfnamefont {P.}~\bibnamefont {Neumann}}, \bibinfo
  {author} {\bibfnamefont {A.}~\bibnamefont {Aird}}, \bibinfo {author}
  {\bibfnamefont {B.}~\bibnamefont {Naydenov}}, \bibinfo {author}
  {\bibfnamefont {G.}~\bibnamefont {Balasubramanian}}, \bibinfo {author}
  {\bibfnamefont {F.}~\bibnamefont {Jelezko}}, \ and\ \bibinfo {author}
  {\bibfnamefont {J.}~\bibnamefont {Wrachtrup}},\ }\href@noop {} {\bibfield
  {journal} {\bibinfo  {journal} {Review of scientific instruments}\ }\textbf
  {\bibinfo {volume} {81}},\ \bibinfo {pages} {043705} (\bibinfo {year}
  {2010})}\BibitemShut {NoStop}%
\bibitem [{\citenamefont {Maertz}\ \emph {et~al.}(2010)\citenamefont {Maertz},
  \citenamefont {Wijnheijmer}, \citenamefont {Fuchs}, \citenamefont
  {Nowakowski},\ and\ \citenamefont {Awschalom}}]{maertz2010vector}%
  \BibitemOpen
  \bibfield  {author} {\bibinfo {author} {\bibfnamefont {B.}~\bibnamefont
  {Maertz}}, \bibinfo {author} {\bibfnamefont {A.}~\bibnamefont {Wijnheijmer}},
  \bibinfo {author} {\bibfnamefont {G.}~\bibnamefont {Fuchs}}, \bibinfo
  {author} {\bibfnamefont {M.}~\bibnamefont {Nowakowski}}, \ and\ \bibinfo
  {author} {\bibfnamefont {D.}~\bibnamefont {Awschalom}},\ }\href@noop {}
  {\bibfield  {journal} {\bibinfo  {journal} {Applied Physics Letters}\
  }\textbf {\bibinfo {volume} {96}},\ \bibinfo {pages} {092504} (\bibinfo
  {year} {2010})}\BibitemShut {NoStop}%
\bibitem [{\citenamefont {Ariyaratne}\ \emph {et~al.}(2018)\citenamefont
  {Ariyaratne}, \citenamefont {Bluvstein}, \citenamefont {Myers},\ and\
  \citenamefont {Jayich}}]{ariyaratne2018nanoscale}%
  \BibitemOpen
  \bibfield  {author} {\bibinfo {author} {\bibfnamefont {A.}~\bibnamefont
  {Ariyaratne}}, \bibinfo {author} {\bibfnamefont {D.}~\bibnamefont
  {Bluvstein}}, \bibinfo {author} {\bibfnamefont {B.~A.}\ \bibnamefont
  {Myers}}, \ and\ \bibinfo {author} {\bibfnamefont {A.~C.~B.}\ \bibnamefont
  {Jayich}},\ }\href@noop {} {\bibfield  {journal} {\bibinfo  {journal} {Nature
  communications}\ }\textbf {\bibinfo {volume} {9}},\ \bibinfo {pages} {1}
  (\bibinfo {year} {2018})}\BibitemShut {NoStop}%
\bibitem [{\citenamefont {Thiel}\ \emph {et~al.}(2016)\citenamefont {Thiel},
  \citenamefont {Rohner}, \citenamefont {Ganzhorn}, \citenamefont {Appel},
  \citenamefont {Neu}, \citenamefont {M{\"u}ller}, \citenamefont {Kleiner},
  \citenamefont {Koelle},\ and\ \citenamefont
  {Maletinsky}}]{thiel2016quantitative}%
  \BibitemOpen
  \bibfield  {author} {\bibinfo {author} {\bibfnamefont {L.}~\bibnamefont
  {Thiel}}, \bibinfo {author} {\bibfnamefont {D.}~\bibnamefont {Rohner}},
  \bibinfo {author} {\bibfnamefont {M.}~\bibnamefont {Ganzhorn}}, \bibinfo
  {author} {\bibfnamefont {P.}~\bibnamefont {Appel}}, \bibinfo {author}
  {\bibfnamefont {E.}~\bibnamefont {Neu}}, \bibinfo {author} {\bibfnamefont
  {B.}~\bibnamefont {M{\"u}ller}}, \bibinfo {author} {\bibfnamefont
  {R.}~\bibnamefont {Kleiner}}, \bibinfo {author} {\bibfnamefont
  {D.}~\bibnamefont {Koelle}}, \ and\ \bibinfo {author} {\bibfnamefont
  {P.}~\bibnamefont {Maletinsky}},\ }\href@noop {} {\bibfield  {journal}
  {\bibinfo  {journal} {Nature nanotechnology}\ }\textbf {\bibinfo {volume}
  {11}},\ \bibinfo {pages} {677} (\bibinfo {year} {2016})}\BibitemShut
  {NoStop}%
\bibitem [{\citenamefont {Pelliccione}\ \emph {et~al.}(2016)\citenamefont
  {Pelliccione}, \citenamefont {Jenkins}, \citenamefont {Ovartchaiyapong},
  \citenamefont {Reetz}, \citenamefont {Emmanouilidou}, \citenamefont {Ni},\
  and\ \citenamefont {Jayich}}]{pelliccione2016scanned}%
  \BibitemOpen
  \bibfield  {author} {\bibinfo {author} {\bibfnamefont {M.}~\bibnamefont
  {Pelliccione}}, \bibinfo {author} {\bibfnamefont {A.}~\bibnamefont
  {Jenkins}}, \bibinfo {author} {\bibfnamefont {P.}~\bibnamefont
  {Ovartchaiyapong}}, \bibinfo {author} {\bibfnamefont {C.}~\bibnamefont
  {Reetz}}, \bibinfo {author} {\bibfnamefont {E.}~\bibnamefont
  {Emmanouilidou}}, \bibinfo {author} {\bibfnamefont {N.}~\bibnamefont {Ni}}, \
  and\ \bibinfo {author} {\bibfnamefont {A.~C.~B.}\ \bibnamefont {Jayich}},\
  }\href@noop {} {\bibfield  {journal} {\bibinfo  {journal} {Nature
  nanotechnology}\ }\textbf {\bibinfo {volume} {11}},\ \bibinfo {pages} {700}
  (\bibinfo {year} {2016})}\BibitemShut {NoStop}%
\bibitem [{\citenamefont {Appel}\ \emph {et~al.}(2015)\citenamefont {Appel},
  \citenamefont {Ganzhorn}, \citenamefont {Neu},\ and\ \citenamefont
  {Maletinsky}}]{appel2015nanoscale}%
  \BibitemOpen
  \bibfield  {author} {\bibinfo {author} {\bibfnamefont {P.}~\bibnamefont
  {Appel}}, \bibinfo {author} {\bibfnamefont {M.}~\bibnamefont {Ganzhorn}},
  \bibinfo {author} {\bibfnamefont {E.}~\bibnamefont {Neu}}, \ and\ \bibinfo
  {author} {\bibfnamefont {P.}~\bibnamefont {Maletinsky}},\ }\href@noop {}
  {\bibfield  {journal} {\bibinfo  {journal} {New Journal of Physics}\ }\textbf
  {\bibinfo {volume} {17}},\ \bibinfo {pages} {112001} (\bibinfo {year}
  {2015})}\BibitemShut {NoStop}%
\bibitem [{\citenamefont {Tetienne}\ \emph {et~al.}(2014)\citenamefont
  {Tetienne}, \citenamefont {Hingant}, \citenamefont {Kim}, \citenamefont
  {Diez}, \citenamefont {Adam}, \citenamefont {Garcia}, \citenamefont {Roch},
  \citenamefont {Rohart}, \citenamefont {Thiaville}, \citenamefont {Ravelosona}
  \emph {et~al.}}]{tetienne2014nanoscale}%
  \BibitemOpen
  \bibfield  {author} {\bibinfo {author} {\bibfnamefont {J.-P.}\ \bibnamefont
  {Tetienne}}, \bibinfo {author} {\bibfnamefont {T.}~\bibnamefont {Hingant}},
  \bibinfo {author} {\bibfnamefont {J.-V.}\ \bibnamefont {Kim}}, \bibinfo
  {author} {\bibfnamefont {L.~H.}\ \bibnamefont {Diez}}, \bibinfo {author}
  {\bibfnamefont {J.-P.}\ \bibnamefont {Adam}}, \bibinfo {author}
  {\bibfnamefont {K.}~\bibnamefont {Garcia}}, \bibinfo {author} {\bibfnamefont
  {J.-F.}\ \bibnamefont {Roch}}, \bibinfo {author} {\bibfnamefont
  {S.}~\bibnamefont {Rohart}}, \bibinfo {author} {\bibfnamefont
  {A.}~\bibnamefont {Thiaville}}, \bibinfo {author} {\bibfnamefont
  {D.}~\bibnamefont {Ravelosona}},  \emph {et~al.},\ }\href@noop {} {\bibfield
  {journal} {\bibinfo  {journal} {Science}\ }\textbf {\bibinfo {volume}
  {344}},\ \bibinfo {pages} {1366} (\bibinfo {year} {2014})}\BibitemShut
  {NoStop}%
\bibitem [{\citenamefont {Grinolds}\ \emph {et~al.}(2013)\citenamefont
  {Grinolds}, \citenamefont {Hong}, \citenamefont {Maletinsky}, \citenamefont
  {Luan}, \citenamefont {Lukin}, \citenamefont {Walsworth},\ and\ \citenamefont
  {Yacoby}}]{grinolds2013nanoscale}%
  \BibitemOpen
  \bibfield  {author} {\bibinfo {author} {\bibfnamefont {M.~S.}\ \bibnamefont
  {Grinolds}}, \bibinfo {author} {\bibfnamefont {S.}~\bibnamefont {Hong}},
  \bibinfo {author} {\bibfnamefont {P.}~\bibnamefont {Maletinsky}}, \bibinfo
  {author} {\bibfnamefont {L.}~\bibnamefont {Luan}}, \bibinfo {author}
  {\bibfnamefont {M.~D.}\ \bibnamefont {Lukin}}, \bibinfo {author}
  {\bibfnamefont {R.~L.}\ \bibnamefont {Walsworth}}, \ and\ \bibinfo {author}
  {\bibfnamefont {A.}~\bibnamefont {Yacoby}},\ }\href@noop {} {\bibfield
  {journal} {\bibinfo  {journal} {Nature Physics}\ }\textbf {\bibinfo {volume}
  {9}},\ \bibinfo {pages} {215} (\bibinfo {year} {2013})}\BibitemShut {NoStop}%
\bibitem [{\citenamefont {Maletinsky}\ \emph {et~al.}(2012)\citenamefont
  {Maletinsky}, \citenamefont {Hong}, \citenamefont {Grinolds}, \citenamefont
  {Hausmann}, \citenamefont {Lukin}, \citenamefont {Walsworth}, \citenamefont
  {Loncar},\ and\ \citenamefont {Yacoby}}]{maletinsky2012robust}%
  \BibitemOpen
  \bibfield  {author} {\bibinfo {author} {\bibfnamefont {P.}~\bibnamefont
  {Maletinsky}}, \bibinfo {author} {\bibfnamefont {S.}~\bibnamefont {Hong}},
  \bibinfo {author} {\bibfnamefont {M.~S.}\ \bibnamefont {Grinolds}}, \bibinfo
  {author} {\bibfnamefont {B.}~\bibnamefont {Hausmann}}, \bibinfo {author}
  {\bibfnamefont {M.~D.}\ \bibnamefont {Lukin}}, \bibinfo {author}
  {\bibfnamefont {R.~L.}\ \bibnamefont {Walsworth}}, \bibinfo {author}
  {\bibfnamefont {M.}~\bibnamefont {Loncar}}, \ and\ \bibinfo {author}
  {\bibfnamefont {A.}~\bibnamefont {Yacoby}},\ }\href@noop {} {\bibfield
  {journal} {\bibinfo  {journal} {Nature nanotechnology}\ }\textbf {\bibinfo
  {volume} {7}},\ \bibinfo {pages} {320} (\bibinfo {year} {2012})}\BibitemShut
  {NoStop}%
\bibitem [{\citenamefont {Rondin}\ \emph {et~al.}(2012)\citenamefont {Rondin},
  \citenamefont {Tetienne}, \citenamefont {Spinicelli}, \citenamefont
  {Dal~Savio}, \citenamefont {Karrai}, \citenamefont {Dantelle}, \citenamefont
  {Thiaville}, \citenamefont {Rohart}, \citenamefont {Roch},\ and\
  \citenamefont {Jacques}}]{rondin2012nanoscale}%
  \BibitemOpen
  \bibfield  {author} {\bibinfo {author} {\bibfnamefont {L.}~\bibnamefont
  {Rondin}}, \bibinfo {author} {\bibfnamefont {J.-P.}\ \bibnamefont
  {Tetienne}}, \bibinfo {author} {\bibfnamefont {P.}~\bibnamefont
  {Spinicelli}}, \bibinfo {author} {\bibfnamefont {C.}~\bibnamefont
  {Dal~Savio}}, \bibinfo {author} {\bibfnamefont {K.}~\bibnamefont {Karrai}},
  \bibinfo {author} {\bibfnamefont {G.}~\bibnamefont {Dantelle}}, \bibinfo
  {author} {\bibfnamefont {A.}~\bibnamefont {Thiaville}}, \bibinfo {author}
  {\bibfnamefont {S.}~\bibnamefont {Rohart}}, \bibinfo {author} {\bibfnamefont
  {J.-F.}\ \bibnamefont {Roch}}, \ and\ \bibinfo {author} {\bibfnamefont
  {V.}~\bibnamefont {Jacques}},\ }\href@noop {} {\bibfield  {journal} {\bibinfo
   {journal} {Applied Physics Letters}\ }\textbf {\bibinfo {volume} {100}},\
  \bibinfo {pages} {153118} (\bibinfo {year} {2012})}\BibitemShut {NoStop}%
\bibitem [{\citenamefont {Li-Li}\ \emph {et~al.}(2010)\citenamefont {Li-Li},
  \citenamefont {Qang-Qin}, \citenamefont {Xin-Yu},\ and\ \citenamefont
  {Dong-Min}}]{li2010design}%
  \BibitemOpen
  \bibfield  {author} {\bibinfo {author} {\bibfnamefont {Y.}~\bibnamefont
  {Li-Li}}, \bibinfo {author} {\bibfnamefont {L.}~\bibnamefont {Qang-Qin}},
  \bibinfo {author} {\bibfnamefont {P.}~\bibnamefont {Xin-Yu}}, \ and\ \bibinfo
  {author} {\bibfnamefont {C.}~\bibnamefont {Dong-Min}},\ }\href@noop {}
  {\bibfield  {journal} {\bibinfo  {journal} {Chinese Physics Letters}\
  }\textbf {\bibinfo {volume} {27}},\ \bibinfo {pages} {038401} (\bibinfo
  {year} {2010})}\BibitemShut {NoStop}%
\bibitem [{\citenamefont {Rudnicki}\ \emph {et~al.}(2013)\citenamefont
  {Rudnicki}, \citenamefont {Mr{\'o}zek}, \citenamefont {M{\l}ynarczyk},\ and\
  \citenamefont {Gawlik}}]{rudnicki2013microwave}%
  \BibitemOpen
  \bibfield  {author} {\bibinfo {author} {\bibfnamefont {D.}~\bibnamefont
  {Rudnicki}}, \bibinfo {author} {\bibfnamefont {M.}~\bibnamefont
  {Mr{\'o}zek}}, \bibinfo {author} {\bibfnamefont {J.}~\bibnamefont
  {M{\l}ynarczyk}}, \ and\ \bibinfo {author} {\bibfnamefont {W.}~\bibnamefont
  {Gawlik}},\ }\href@noop {} {\bibfield  {journal} {\bibinfo  {journal}
  {Photonics Letters of Poland}\ }\textbf {\bibinfo {volume} {5}},\ \bibinfo
  {pages} {143} (\bibinfo {year} {2013})}\BibitemShut {NoStop}%
\bibitem [{\citenamefont {Mr{\'o}zek}\ \emph {et~al.}(2015)\citenamefont
  {Mr{\'o}zek}, \citenamefont {Mlynarczyk}, \citenamefont {Rudnicki},\ and\
  \citenamefont {Gawlik}}]{mrozek2015circularly}%
  \BibitemOpen
  \bibfield  {author} {\bibinfo {author} {\bibfnamefont {M.}~\bibnamefont
  {Mr{\'o}zek}}, \bibinfo {author} {\bibfnamefont {J.}~\bibnamefont
  {Mlynarczyk}}, \bibinfo {author} {\bibfnamefont {D.~S.}\ \bibnamefont
  {Rudnicki}}, \ and\ \bibinfo {author} {\bibfnamefont {W.}~\bibnamefont
  {Gawlik}},\ }\href@noop {} {\bibfield  {journal} {\bibinfo  {journal}
  {Applied Physics Letters}\ }\textbf {\bibinfo {volume} {107}},\ \bibinfo
  {pages} {013505} (\bibinfo {year} {2015})}\BibitemShut {NoStop}%
\bibitem [{\citenamefont {Sasaki}\ \emph {et~al.}(2016)\citenamefont {Sasaki},
  \citenamefont {Monnai}, \citenamefont {Saijo}, \citenamefont {Fujita},
  \citenamefont {Watanabe}, \citenamefont {Ishi-Hayase}, \citenamefont {Itoh},\
  and\ \citenamefont {Abe}}]{sasaki2016broadband}%
  \BibitemOpen
  \bibfield  {author} {\bibinfo {author} {\bibfnamefont {K.}~\bibnamefont
  {Sasaki}}, \bibinfo {author} {\bibfnamefont {Y.}~\bibnamefont {Monnai}},
  \bibinfo {author} {\bibfnamefont {S.}~\bibnamefont {Saijo}}, \bibinfo
  {author} {\bibfnamefont {R.}~\bibnamefont {Fujita}}, \bibinfo {author}
  {\bibfnamefont {H.}~\bibnamefont {Watanabe}}, \bibinfo {author}
  {\bibfnamefont {J.}~\bibnamefont {Ishi-Hayase}}, \bibinfo {author}
  {\bibfnamefont {K.~M.}\ \bibnamefont {Itoh}}, \ and\ \bibinfo {author}
  {\bibfnamefont {E.}~\bibnamefont {Abe}},\ }\href@noop {} {\bibfield
  {journal} {\bibinfo  {journal} {Review of Scientific Instruments}\ }\textbf
  {\bibinfo {volume} {87}},\ \bibinfo {pages} {053904} (\bibinfo {year}
  {2016})}\BibitemShut {NoStop}%
\bibitem [{\citenamefont {Qin}\ \emph {et~al.}(2018)\citenamefont {Qin},
  \citenamefont {Fu}, \citenamefont {Zhang}, \citenamefont {Zhao},
  \citenamefont {Gao}, \citenamefont {Yuan}, \citenamefont {Ma}, \citenamefont
  {Shi},\ and\ \citenamefont {Liu}}]{qin2018near}%
  \BibitemOpen
  \bibfield  {author} {\bibinfo {author} {\bibfnamefont {L.}~\bibnamefont
  {Qin}}, \bibinfo {author} {\bibfnamefont {Y.}~\bibnamefont {Fu}}, \bibinfo
  {author} {\bibfnamefont {S.}~\bibnamefont {Zhang}}, \bibinfo {author}
  {\bibfnamefont {J.}~\bibnamefont {Zhao}}, \bibinfo {author} {\bibfnamefont
  {J.}~\bibnamefont {Gao}}, \bibinfo {author} {\bibfnamefont {H.}~\bibnamefont
  {Yuan}}, \bibinfo {author} {\bibfnamefont {Z.}~\bibnamefont {Ma}}, \bibinfo
  {author} {\bibfnamefont {Y.}~\bibnamefont {Shi}}, \ and\ \bibinfo {author}
  {\bibfnamefont {J.}~\bibnamefont {Liu}},\ }\href@noop {} {\bibfield
  {journal} {\bibinfo  {journal} {Japanese Journal of Applied Physics}\
  }\textbf {\bibinfo {volume} {57}},\ \bibinfo {pages} {072201} (\bibinfo
  {year} {2018})}\BibitemShut {NoStop}%
\bibitem [{\citenamefont {Chen}\ \emph {et~al.}(2018)\citenamefont {Chen},
  \citenamefont {Guo}, \citenamefont {Li}, \citenamefont {Wu}, \citenamefont
  {Zhu}, \citenamefont {Zhao}, \citenamefont {Wang}, \citenamefont {Zhang},
  \citenamefont {Zhao}, \citenamefont {Liu} \emph {et~al.}}]{chen2018large}%
  \BibitemOpen
  \bibfield  {author} {\bibinfo {author} {\bibfnamefont {Y.}~\bibnamefont
  {Chen}}, \bibinfo {author} {\bibfnamefont {H.}~\bibnamefont {Guo}}, \bibinfo
  {author} {\bibfnamefont {W.}~\bibnamefont {Li}}, \bibinfo {author}
  {\bibfnamefont {D.}~\bibnamefont {Wu}}, \bibinfo {author} {\bibfnamefont
  {Q.}~\bibnamefont {Zhu}}, \bibinfo {author} {\bibfnamefont {B.}~\bibnamefont
  {Zhao}}, \bibinfo {author} {\bibfnamefont {L.}~\bibnamefont {Wang}}, \bibinfo
  {author} {\bibfnamefont {Y.}~\bibnamefont {Zhang}}, \bibinfo {author}
  {\bibfnamefont {R.}~\bibnamefont {Zhao}}, \bibinfo {author} {\bibfnamefont
  {W.}~\bibnamefont {Liu}},  \emph {et~al.},\ }\href@noop {} {\bibfield
  {journal} {\bibinfo  {journal} {Applied Physics Express}\ }\textbf {\bibinfo
  {volume} {11}},\ \bibinfo {pages} {123001} (\bibinfo {year}
  {2018})}\BibitemShut {NoStop}%
\bibitem [{\citenamefont {Dong}\ \emph {et~al.}(2018)\citenamefont {Dong},
  \citenamefont {Hu}, \citenamefont {Liu}, \citenamefont {Yang}, \citenamefont
  {Wang},\ and\ \citenamefont {Du}}]{dong2018fiber}%
  \BibitemOpen
  \bibfield  {author} {\bibinfo {author} {\bibfnamefont {M.}~\bibnamefont
  {Dong}}, \bibinfo {author} {\bibfnamefont {Z.}~\bibnamefont {Hu}}, \bibinfo
  {author} {\bibfnamefont {Y.}~\bibnamefont {Liu}}, \bibinfo {author}
  {\bibfnamefont {B.}~\bibnamefont {Yang}}, \bibinfo {author} {\bibfnamefont
  {Y.}~\bibnamefont {Wang}}, \ and\ \bibinfo {author} {\bibfnamefont
  {G.}~\bibnamefont {Du}},\ }\href@noop {} {\bibfield  {journal} {\bibinfo
  {journal} {Applied Physics Letters}\ }\textbf {\bibinfo {volume} {113}},\
  \bibinfo {pages} {131105} (\bibinfo {year} {2018})}\BibitemShut {NoStop}%
\bibitem [{\citenamefont {Soshenko}\ \emph {et~al.}(2018)\citenamefont
  {Soshenko}, \citenamefont {Rubinas}, \citenamefont {Vorobyov}, \citenamefont
  {Bolshedvorskii}, \citenamefont {Kapitanova}, \citenamefont {Sorokin},\ and\
  \citenamefont {Akimov}}]{soshenko2018microwave}%
  \BibitemOpen
  \bibfield  {author} {\bibinfo {author} {\bibfnamefont {V.}~\bibnamefont
  {Soshenko}}, \bibinfo {author} {\bibfnamefont {O.}~\bibnamefont {Rubinas}},
  \bibinfo {author} {\bibfnamefont {V.}~\bibnamefont {Vorobyov}}, \bibinfo
  {author} {\bibfnamefont {S.}~\bibnamefont {Bolshedvorskii}}, \bibinfo
  {author} {\bibfnamefont {P.}~\bibnamefont {Kapitanova}}, \bibinfo {author}
  {\bibfnamefont {V.}~\bibnamefont {Sorokin}}, \ and\ \bibinfo {author}
  {\bibfnamefont {A.}~\bibnamefont {Akimov}},\ }\href@noop {} {\bibfield
  {journal} {\bibinfo  {journal} {Bulletin of the Lebedev Physics Institute}\
  }\textbf {\bibinfo {volume} {45}},\ \bibinfo {pages} {237} (\bibinfo {year}
  {2018})}\BibitemShut {NoStop}%
\bibitem [{\citenamefont {Alegre}\ \emph {et~al.}(2007)\citenamefont {Alegre},
  \citenamefont {Santori}, \citenamefont {Medeiros-Ribeiro},\ and\
  \citenamefont {Beausoleil}}]{alegre2007polarization}%
  \BibitemOpen
  \bibfield  {author} {\bibinfo {author} {\bibfnamefont {T.~P.~M.}\
  \bibnamefont {Alegre}}, \bibinfo {author} {\bibfnamefont {C.}~\bibnamefont
  {Santori}}, \bibinfo {author} {\bibfnamefont {G.}~\bibnamefont
  {Medeiros-Ribeiro}}, \ and\ \bibinfo {author} {\bibfnamefont {R.~G.}\
  \bibnamefont {Beausoleil}},\ }\href@noop {} {\bibfield  {journal} {\bibinfo
  {journal} {Physical Review B}\ }\textbf {\bibinfo {volume} {76}},\ \bibinfo
  {pages} {165205} (\bibinfo {year} {2007})}\BibitemShut {NoStop}%
\bibitem [{\citenamefont {Bayat}\ \emph {et~al.}(2014)\citenamefont {Bayat},
  \citenamefont {Choy}, \citenamefont {Farrokh~Baroughi}, \citenamefont
  {Meesala},\ and\ \citenamefont {Loncar}}]{bayat2014efficient}%
  \BibitemOpen
  \bibfield  {author} {\bibinfo {author} {\bibfnamefont {K.}~\bibnamefont
  {Bayat}}, \bibinfo {author} {\bibfnamefont {J.}~\bibnamefont {Choy}},
  \bibinfo {author} {\bibfnamefont {M.}~\bibnamefont {Farrokh~Baroughi}},
  \bibinfo {author} {\bibfnamefont {S.}~\bibnamefont {Meesala}}, \ and\
  \bibinfo {author} {\bibfnamefont {M.}~\bibnamefont {Loncar}},\ }\href@noop {}
  {\bibfield  {journal} {\bibinfo  {journal} {Nano letters}\ }\textbf {\bibinfo
  {volume} {14}},\ \bibinfo {pages} {1208} (\bibinfo {year}
  {2014})}\BibitemShut {NoStop}%
\bibitem [{\citenamefont {Herrmann}\ \emph {et~al.}(2016)\citenamefont
  {Herrmann}, \citenamefont {Appleton}, \citenamefont {Sasaki}, \citenamefont
  {Monnai}, \citenamefont {Teraji}, \citenamefont {Itoh},\ and\ \citenamefont
  {Abe}}]{herrmann2016polarization}%
  \BibitemOpen
  \bibfield  {author} {\bibinfo {author} {\bibfnamefont {J.}~\bibnamefont
  {Herrmann}}, \bibinfo {author} {\bibfnamefont {M.~A.}\ \bibnamefont
  {Appleton}}, \bibinfo {author} {\bibfnamefont {K.}~\bibnamefont {Sasaki}},
  \bibinfo {author} {\bibfnamefont {Y.}~\bibnamefont {Monnai}}, \bibinfo
  {author} {\bibfnamefont {T.}~\bibnamefont {Teraji}}, \bibinfo {author}
  {\bibfnamefont {K.~M.}\ \bibnamefont {Itoh}}, \ and\ \bibinfo {author}
  {\bibfnamefont {E.}~\bibnamefont {Abe}},\ }\href@noop {} {\bibfield
  {journal} {\bibinfo  {journal} {Applied Physics Letters}\ }\textbf {\bibinfo
  {volume} {109}},\ \bibinfo {pages} {183111} (\bibinfo {year}
  {2016})}\BibitemShut {NoStop}%
\bibitem [{\citenamefont {Zhang}\ \emph {et~al.}(2016)\citenamefont {Zhang},
  \citenamefont {Zhang}, \citenamefont {Xu}, \citenamefont {Ding},
  \citenamefont {Quan}, \citenamefont {Tang},\ and\ \citenamefont
  {Yuan}}]{zhang2016microwave}%
  \BibitemOpen
  \bibfield  {author} {\bibinfo {author} {\bibfnamefont {N.}~\bibnamefont
  {Zhang}}, \bibinfo {author} {\bibfnamefont {C.}~\bibnamefont {Zhang}},
  \bibinfo {author} {\bibfnamefont {L.}~\bibnamefont {Xu}}, \bibinfo {author}
  {\bibfnamefont {M.}~\bibnamefont {Ding}}, \bibinfo {author} {\bibfnamefont
  {W.}~\bibnamefont {Quan}}, \bibinfo {author} {\bibfnamefont {Z.}~\bibnamefont
  {Tang}}, \ and\ \bibinfo {author} {\bibfnamefont {H.}~\bibnamefont {Yuan}},\
  }\href@noop {} {\bibfield  {journal} {\bibinfo  {journal} {Applied Magnetic
  Resonance}\ }\textbf {\bibinfo {volume} {47}},\ \bibinfo {pages} {589}
  (\bibinfo {year} {2016})}\BibitemShut {NoStop}%
\bibitem [{\citenamefont {Yang}\ \emph {et~al.}(2019)\citenamefont {Yang},
  \citenamefont {Zhang}, \citenamefont {Yuan}, \citenamefont {Bian},
  \citenamefont {Fan},\ and\ \citenamefont {Li}}]{yang2019microstrip}%
  \BibitemOpen
  \bibfield  {author} {\bibinfo {author} {\bibfnamefont {X.}~\bibnamefont
  {Yang}}, \bibinfo {author} {\bibfnamefont {N.}~\bibnamefont {Zhang}},
  \bibinfo {author} {\bibfnamefont {H.}~\bibnamefont {Yuan}}, \bibinfo {author}
  {\bibfnamefont {G.}~\bibnamefont {Bian}}, \bibinfo {author} {\bibfnamefont
  {P.}~\bibnamefont {Fan}}, \ and\ \bibinfo {author} {\bibfnamefont
  {M.}~\bibnamefont {Li}},\ }\href@noop {} {\bibfield  {journal} {\bibinfo
  {journal} {AIP Advances}\ }\textbf {\bibinfo {volume} {9}},\ \bibinfo {pages}
  {075213} (\bibinfo {year} {2019})}\BibitemShut {NoStop}%
\bibitem [{\citenamefont {Mariani}\ \emph {et~al.}(2020)\citenamefont
  {Mariani}, \citenamefont {Nomoto}, \citenamefont {Kashiwaya},\ and\
  \citenamefont {Nomura}}]{mariani2020system}%
  \BibitemOpen
  \bibfield  {author} {\bibinfo {author} {\bibfnamefont {G.}~\bibnamefont
  {Mariani}}, \bibinfo {author} {\bibfnamefont {S.}~\bibnamefont {Nomoto}},
  \bibinfo {author} {\bibfnamefont {S.}~\bibnamefont {Kashiwaya}}, \ and\
  \bibinfo {author} {\bibfnamefont {S.}~\bibnamefont {Nomura}},\ }\href@noop {}
  {\bibfield  {journal} {\bibinfo  {journal} {Scientific Reports}\ }\textbf
  {\bibinfo {volume} {10}},\ \bibinfo {pages} {1} (\bibinfo {year}
  {2020})}\BibitemShut {NoStop}%
\bibitem [{\citenamefont {Yaroshenko}\ \emph {et~al.}(2020)\citenamefont
  {Yaroshenko}, \citenamefont {Soshenko}, \citenamefont {Vorobyov},
  \citenamefont {Bolshedvorskii}, \citenamefont {Nenasheva}, \citenamefont
  {Kotel’nikov}, \citenamefont {Akimov},\ and\ \citenamefont
  {Kapitanova}}]{yaroshenko2020circularly}%
  \BibitemOpen
  \bibfield  {author} {\bibinfo {author} {\bibfnamefont {V.}~\bibnamefont
  {Yaroshenko}}, \bibinfo {author} {\bibfnamefont {V.}~\bibnamefont
  {Soshenko}}, \bibinfo {author} {\bibfnamefont {V.}~\bibnamefont {Vorobyov}},
  \bibinfo {author} {\bibfnamefont {S.}~\bibnamefont {Bolshedvorskii}},
  \bibinfo {author} {\bibfnamefont {E.}~\bibnamefont {Nenasheva}}, \bibinfo
  {author} {\bibfnamefont {I.}~\bibnamefont {Kotel’nikov}}, \bibinfo {author}
  {\bibfnamefont {A.}~\bibnamefont {Akimov}}, \ and\ \bibinfo {author}
  {\bibfnamefont {P.}~\bibnamefont {Kapitanova}},\ }\href@noop {} {\bibfield
  {journal} {\bibinfo  {journal} {Review of Scientific Instruments}\ }\textbf
  {\bibinfo {volume} {91}},\ \bibinfo {pages} {035003} (\bibinfo {year}
  {2020})}\BibitemShut {NoStop}%
\bibitem [{\citenamefont {Horowitz}\ \emph {et~al.}(2012)\citenamefont
  {Horowitz}, \citenamefont {Alem{\'a}n}, \citenamefont {Christle},
  \citenamefont {Cleland},\ and\ \citenamefont
  {Awschalom}}]{horowitz2012electron}%
  \BibitemOpen
  \bibfield  {author} {\bibinfo {author} {\bibfnamefont {V.~R.}\ \bibnamefont
  {Horowitz}}, \bibinfo {author} {\bibfnamefont {B.~J.}\ \bibnamefont
  {Alem{\'a}n}}, \bibinfo {author} {\bibfnamefont {D.~J.}\ \bibnamefont
  {Christle}}, \bibinfo {author} {\bibfnamefont {A.~N.}\ \bibnamefont
  {Cleland}}, \ and\ \bibinfo {author} {\bibfnamefont {D.~D.}\ \bibnamefont
  {Awschalom}},\ }\href@noop {} {\bibfield  {journal} {\bibinfo  {journal}
  {Proceedings of the National Academy of Sciences}\ }\textbf {\bibinfo
  {volume} {109}},\ \bibinfo {pages} {13493} (\bibinfo {year}
  {2012})}\BibitemShut {NoStop}%
\bibitem [{\citenamefont {Weiland}(1977)}]{weiland1977discretization}%
  \BibitemOpen
  \bibfield  {author} {\bibinfo {author} {\bibfnamefont {T.}~\bibnamefont
  {Weiland}},\ }\href@noop {} {\bibfield  {journal} {\bibinfo  {journal}
  {Archiv Elektronik und Uebertragungstechnik}\ }\textbf {\bibinfo {volume}
  {31}},\ \bibinfo {pages} {116} (\bibinfo {year} {1977})}\BibitemShut
  {NoStop}%
\bibitem [{\citenamefont {Jelezko}\ \emph {et~al.}(2004)\citenamefont
  {Jelezko}, \citenamefont {Gaebel}, \citenamefont {Popa}, \citenamefont
  {Gruber},\ and\ \citenamefont {Wrachtrup}}]{Jelezko2004}%
  \BibitemOpen
  \bibfield  {author} {\bibinfo {author} {\bibfnamefont {F.}~\bibnamefont
  {Jelezko}}, \bibinfo {author} {\bibfnamefont {T.}~\bibnamefont {Gaebel}},
  \bibinfo {author} {\bibfnamefont {I.}~\bibnamefont {Popa}}, \bibinfo {author}
  {\bibfnamefont {A.}~\bibnamefont {Gruber}}, \ and\ \bibinfo {author}
  {\bibfnamefont {J.}~\bibnamefont {Wrachtrup}},\ }\href {\doibase
  10.1103/PhysRevLett.92.076401} {\bibfield  {journal} {\bibinfo  {journal}
  {Phys. Rev. Lett.}\ }\textbf {\bibinfo {volume} {92}},\ \bibinfo {pages}
  {076401} (\bibinfo {year} {2004})}\BibitemShut {NoStop}%
\bibitem [{\citenamefont {Gruber}\ \emph {et~al.}(1997)\citenamefont {Gruber},
  \citenamefont {Dr{\"a}benstedt}, \citenamefont {Tietz}, \citenamefont
  {Fleury}, \citenamefont {Wrachtrup},\ and\ \citenamefont
  {Borczyskowski}}]{Gruber2012}%
  \BibitemOpen
  \bibfield  {author} {\bibinfo {author} {\bibfnamefont {A.}~\bibnamefont
  {Gruber}}, \bibinfo {author} {\bibfnamefont {A.}~\bibnamefont
  {Dr{\"a}benstedt}}, \bibinfo {author} {\bibfnamefont {C.}~\bibnamefont
  {Tietz}}, \bibinfo {author} {\bibfnamefont {L.}~\bibnamefont {Fleury}},
  \bibinfo {author} {\bibfnamefont {J.}~\bibnamefont {Wrachtrup}}, \ and\
  \bibinfo {author} {\bibfnamefont {C.~v.}\ \bibnamefont {Borczyskowski}},\
  }\href {\doibase 10.1126/science.276.5321.2012} {\bibfield  {journal}
  {\bibinfo  {journal} {Science}\ }\textbf {\bibinfo {volume} {276}},\ \bibinfo
  {pages} {2012} (\bibinfo {year} {1997})}\BibitemShut {NoStop}%
\bibitem [{\citenamefont {MacQuarrie}\ \emph {et~al.}(2015)\citenamefont
  {MacQuarrie}, \citenamefont {Gosavi}, \citenamefont {Bhave},\ and\
  \citenamefont {Fuchs}}]{MacQuarrie2015}%
  \BibitemOpen
  \bibfield  {author} {\bibinfo {author} {\bibfnamefont {E.~R.}\ \bibnamefont
  {MacQuarrie}}, \bibinfo {author} {\bibfnamefont {T.~A.}\ \bibnamefont
  {Gosavi}}, \bibinfo {author} {\bibfnamefont {S.~A.}\ \bibnamefont {Bhave}}, \
  and\ \bibinfo {author} {\bibfnamefont {G.~D.}\ \bibnamefont {Fuchs}},\ }\href
  {\doibase 10.1103/PhysRevB.92.224419} {\bibfield  {journal} {\bibinfo
  {journal} {Phys. Rev. B}\ }\textbf {\bibinfo {volume} {92}},\ \bibinfo
  {pages} {224419} (\bibinfo {year} {2015})}\BibitemShut {NoStop}%
\bibitem [{\citenamefont {Mizuno}\ \emph {et~al.}(2021)\citenamefont {Mizuno},
  \citenamefont {Nakajima}, \citenamefont {Ishiwata}, \citenamefont {Hatano},\
  and\ \citenamefont {Iwasaki}}]{mizuno2021electron}%
  \BibitemOpen
  \bibfield  {author} {\bibinfo {author} {\bibfnamefont {K.}~\bibnamefont
  {Mizuno}}, \bibinfo {author} {\bibfnamefont {M.}~\bibnamefont {Nakajima}},
  \bibinfo {author} {\bibfnamefont {H.}~\bibnamefont {Ishiwata}}, \bibinfo
  {author} {\bibfnamefont {M.}~\bibnamefont {Hatano}}, \ and\ \bibinfo {author}
  {\bibfnamefont {T.}~\bibnamefont {Iwasaki}},\ }\href@noop {} {\bibfield
  {journal} {\bibinfo  {journal} {Applied Physics Express}\ }\textbf {\bibinfo
  {volume} {14}},\ \bibinfo {pages} {032001} (\bibinfo {year}
  {2021})}\BibitemShut {NoStop}%
\bibitem [{\citenamefont {Osterkamp}\ \emph {et~al.}(2019)\citenamefont
  {Osterkamp}, \citenamefont {Mangold}, \citenamefont {Lang}, \citenamefont
  {Balasubramanian}, \citenamefont {Teraji}, \citenamefont {Naydenov},\ and\
  \citenamefont {Jelezko}}]{osterkamp2019engineering}%
  \BibitemOpen
  \bibfield  {author} {\bibinfo {author} {\bibfnamefont {C.}~\bibnamefont
  {Osterkamp}}, \bibinfo {author} {\bibfnamefont {M.}~\bibnamefont {Mangold}},
  \bibinfo {author} {\bibfnamefont {J.}~\bibnamefont {Lang}}, \bibinfo {author}
  {\bibfnamefont {P.}~\bibnamefont {Balasubramanian}}, \bibinfo {author}
  {\bibfnamefont {T.}~\bibnamefont {Teraji}}, \bibinfo {author} {\bibfnamefont
  {B.}~\bibnamefont {Naydenov}}, \ and\ \bibinfo {author} {\bibfnamefont
  {F.}~\bibnamefont {Jelezko}},\ }\href@noop {} {\bibfield  {journal} {\bibinfo
   {journal} {Scientific reports}\ }\textbf {\bibinfo {volume} {9}},\ \bibinfo
  {pages} {1} (\bibinfo {year} {2019})}\BibitemShut {NoStop}%
\bibitem [{\citenamefont {Dr{\'e}au}\ \emph {et~al.}(2011)\citenamefont
  {Dr{\'e}au}, \citenamefont {Lesik}, \citenamefont {Rondin}, \citenamefont
  {Spinicelli}, \citenamefont {Arcizet}, \citenamefont {Roch},\ and\
  \citenamefont {Jacques}}]{dreau2011avoiding}%
  \BibitemOpen
  \bibfield  {author} {\bibinfo {author} {\bibfnamefont {A.}~\bibnamefont
  {Dr{\'e}au}}, \bibinfo {author} {\bibfnamefont {M.}~\bibnamefont {Lesik}},
  \bibinfo {author} {\bibfnamefont {L.}~\bibnamefont {Rondin}}, \bibinfo
  {author} {\bibfnamefont {P.}~\bibnamefont {Spinicelli}}, \bibinfo {author}
  {\bibfnamefont {O.}~\bibnamefont {Arcizet}}, \bibinfo {author} {\bibfnamefont
  {J.-F.}\ \bibnamefont {Roch}}, \ and\ \bibinfo {author} {\bibfnamefont
  {V.}~\bibnamefont {Jacques}},\ }\href@noop {} {\bibfield  {journal} {\bibinfo
   {journal} {Physical Review B}\ }\textbf {\bibinfo {volume} {84}},\ \bibinfo
  {pages} {195204} (\bibinfo {year} {2011})}\BibitemShut {NoStop}%
\bibitem [{\citenamefont {Rembold}\ \emph {et~al.}(2020)\citenamefont
  {Rembold}, \citenamefont {Oshnik}, \citenamefont {Müller}, \citenamefont
  {Montangero}, \citenamefont {Calarco},\ and\ \citenamefont
  {Neu}}]{Rembold2020}%
  \BibitemOpen
  \bibfield  {author} {\bibinfo {author} {\bibfnamefont {P.}~\bibnamefont
  {Rembold}}, \bibinfo {author} {\bibfnamefont {N.}~\bibnamefont {Oshnik}},
  \bibinfo {author} {\bibfnamefont {M.~M.}\ \bibnamefont {Müller}}, \bibinfo
  {author} {\bibfnamefont {S.}~\bibnamefont {Montangero}}, \bibinfo {author}
  {\bibfnamefont {T.}~\bibnamefont {Calarco}}, \ and\ \bibinfo {author}
  {\bibfnamefont {E.}~\bibnamefont {Neu}},\ }\href {\doibase 10.1116/5.0006785}
  {\bibfield  {journal} {\bibinfo  {journal} {AVS Quantum Science}\ }\textbf
  {\bibinfo {volume} {2}},\ \bibinfo {pages} {024701} (\bibinfo {year}
  {2020})},\ \Eprint {http://arxiv.org/abs/https://doi.org/10.1116/5.0006785}
  {https://doi.org/10.1116/5.0006785} \BibitemShut {NoStop}%
\end{thebibliography}%

\end{document}